\documentclass{article}

\PassOptionsToPackage{numbers, compress}{natbib}
\usepackage[main,final]{neurips_2025}

\usepackage[utf8]{inputenc}
\usepackage[T1]{fontenc}
\usepackage{url}
\usepackage{booktabs}
\usepackage{amsfonts} 
\usepackage{amsmath}
\usepackage{amssymb}
\usepackage{nicefrac}
\usepackage{xcolor} 
\definecolor{Periwinkle}{RGB}{204,204,255}
\definecolor{mygrey}{gray}{0.6}
\usepackage{xspace}
\usepackage{pifont}
\usepackage{graphicx}
\usepackage{subcaption}
\usepackage{siunitx}
\usepackage{multirow}
\usepackage{enumitem}
\usepackage{makecell}
\usepackage{placeins}
\usepackage{colortbl}
\usepackage{algorithm}
\usepackage{algorithmicx}
\usepackage{algpseudocode}
\usepackage{caption}
\usepackage{tabularx}
\usepackage[table]{xcolor}
\usepackage{fvextra}

\usepackage{listings}
\usepackage{tcolorbox}
\tcbuselibrary{skins,breakable,listings}

\newcommand{\std}[1]{\scriptsize{$\pm$#1}}

\newcommand{\obsbox}[1]{%
    \begin{tcolorbox}[colframe=black!70, colback=lightgray!15, boxrule=1pt, arc=2mm]
        \small#1
    \end{tcolorbox}
}

\usepackage{hyperref}
\hypersetup{
    colorlinks=true,
    linkcolor=red,
    citecolor=cyan,
    filecolor=magenta,      
    urlcolor=magenta,
    }

\usepackage{soul}
\definecolor{c1}{HTML}{f9aba3}   
\definecolor{c2}{HTML}{db8575}    
\definecolor{c3}{HTML}{e9684f}  
\definecolor{c4}{HTML}{96b78c }   
\definecolor{c5}{HTML}{ecd1b4 }   
\definecolor{c6}{HTML}{fdea92}   

\definecolor{c7}{HTML}{a2bbdb}   
\definecolor{c8}{HTML}{333366}   
\definecolor{c9}{HTML}{336633}   
\definecolor{c10}{HTML}{663300}  

\newcommand{\wrc}[1]{{\sethlcolor{c1}\hl{#1}}} 
\newcommand{\ik} [1]{{\sethlcolor{c2}\hl{#1}}} 
\newcommand{\ek} [1]{{\sethlcolor{c3}\hl{#1}}} 
\newcommand{\rf} [1]{{\sethlcolor{c4}\hl{#1}}}  
\newcommand{\wa} [1]{{\sethlcolor{c5}\hl{#1}}} 
\newcommand{\re} [1]{{\sethlcolor{c6}\hl{#1}}} 
\newcommand{\sq} [1]{{\sethlcolor{c7}\hl{#1}}} 

\newcommand{\he}     [1]{{\textcolor{c7}{#1}}} 
\newcommand{\cEight}[1]{{\textcolor{c8}{#1}}}
\newcommand{\cNine} [1]{{\textcolor{c9}{#1}}}
\newcommand{\cTen}  [1]{{\textcolor{c10}{#1}}}

\definecolor{DeepBlue}{HTML}{0d1746}

\usepackage{microtype}

\title{Auditing Meta-Cognitive Hallucinations in \\ Reasoning Large Language Models}

\author{%
\parbox{\linewidth}{\centering
{\bfseries
\makebox[\textwidth][c]{%
\begin{tabular}{@{}c@{\hspace{1.25em}}c@{\hspace{1.25em}}c@{\hspace{1.25em}}c@{}}
Haolang Lu\textsuperscript{1}\thanks{Equal contribution.} &
Yilian Liu\textsuperscript{1}\footnotemark[1] &
Jingxin Xu\textsuperscript{1}\footnotemark[1] &
Guoshun Nan\textsuperscript{1}\thanks{Corresponding author.}
\end{tabular}}\\
\makebox[\textwidth][c]{%
\begin{tabular}{@{}c@{\hspace{1.25em}}c@{\hspace{1.25em}}c@{}}
Yuanlong Yu\textsuperscript{1} &
Zhican Chen\textsuperscript{1} &
Kun Wang\textsuperscript{2}
\end{tabular}}
}
\\[0.7ex]
{\normalfont
\textsuperscript{1}Beijing University of Posts and Telecommunications, China\\[0.8ex]
\textsuperscript{2}Nanyang Technological University, Singapore\\[0.35em]
\texttt{\{luhaolang2507, liuyilian, xujingxin, nanguo2021\}@bupt.edu.cn} \qquad}
}
}

\begin{document}

\maketitle

\begin{abstract}
The development of Reasoning Large Language Models (RLLMs) has significantly improved multi-step reasoning capabilities, but it has also made hallucination problems more frequent and harder to eliminate. While existing approaches address hallucination through external knowledge integration, model parameter analysis, or self-verification mechanisms, they fail to provide a comprehensive insight into how hallucinations \textbf{emerge} and \textbf{evolve} throughout the reasoning chain. In this work, we investigate hallucination causality under constrained knowledge domains by auditing the Chain-of-Thought (CoT) trajectory and assessing the model’s cognitive confidence in potentially erroneous or biased claims.
Analysis reveals that in long-CoT settings, RLLMs may iteratively reinforce biases and errors through flawed reflective processes, ultimately inducing hallucinated reasoning paths.
Counterintuitively, even with interventions at hallucination origins, reasoning chains display pronounced ``chain disloyalty'', resisting correction and sustaining flawed trajectories.
We further point out that existing hallucination detection methods are \textit{less reliable and interpretable than previously assumed}, especially in complex multi-step reasoning contexts.
Unlike circuit tracing that requires access to model parameters, our auditing \textbf{enables more interpretable long-chain hallucination attribution in black-box settings}, demonstrating stronger generalizability and practical utility.
Our code is available at~\href{https://github.com/Winnie-Lian/AHa_Meta_Cognitive}{this link}.
    
\end{abstract}
\section{Introduction}

\vspace{-0.5em}
Reasoning Large Language Models (RLLMs)~\cite{chen2025reasoning,zhong2024evaluation,li2025system} have gained increasing attention for their ability to perform multi-step reasoning through structured Chain-of-Thought (CoT) and self-reflection mechanisms~\cite{sun2023survey,ji2023towards,li2025survey,yu2024natural}.
While these mechanisms improve performance in complex reasoning tasks~\cite{yax2024studying,chen2025reasoning,team2025kimi}, they also exacerbate the risk of hallucination by amplifying early-stage errors across extended reasoning chains.
In particular, hallucinations in long-CoT settings may be iteratively revised, elaborated, or reframed through the reasoning process.
This results in final answers that appear coherent yet embed deeply masked factual errors, while users often focus on the answer rather than the reasoning process, thus failing to recognize the presence of hallucinations~\cite{ashktorab2024emerging,maksimov2024deeppavlov}.

Numerous research institutions and groups have made significant efforts to address hallucination in LLMs~\cite{bang2023multitask,ji2023survey,ji2023towards,zhang2024language}.
\textbf{At the surface level}, existing literature mainly focuses on detection and mitigation methods that leverage external knowledge sources (e.g., knowledge bases)~\cite{min2023factscore,bayat2023fleek}, or utilize self-checking mechanisms~\cite{ji2023towards,gosmar2025hallucination}.
Alternatively, other methods are algorithm-based, such as using perplexity~\cite{guo2022survey,fadeeva2024fact} or detecting the model's hidden states~\cite{sriramanan2024llm,du2024haloscope,cheninside,zhu2024pollmgraph} to identify hallucinations in longer model outputs.
In the context of CoT reasoning, some studies have explored the multi-step reasoning phenomenon inherent to CoT~\cite{jin2024impact,chen2025reasoning,li2025system,yang2024large}, aiming to understand its implications for the reasoning model's output accuracy~\cite{kim2024lachesis} and reliability~\cite{mondorfbeyond,wan2024cot}.
\textbf{At a deeper level}, understanding the underlying mechanisms of hallucination is critical for improving RLLMs, as the complexity of the reasoning chain often means that surface-level detection methods may not guarantee optimal outcomes.
In this regard, works have made notable contributions by leveraging sparse encoders~\cite{abdaljalil2025safe} and causal probing~\cite{zhou2024mitigating} to trace which components of the model contribute to specific outputs~\cite{simhi2024constructing}.

\begin{figure*}[t]
    \centering

    \centering
    \begin{subfigure}[t]{0.33\textwidth}
        \centering
        \includegraphics[width=\textwidth]{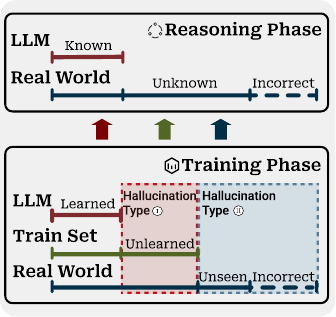}

        \caption{Knowledge Domain}

        \label{fig:Knowledge Domain}
    \end{subfigure}
    \hfill
    \begin{subfigure}[t]{0.32\textwidth}
        \centering
        \includegraphics[width=\textwidth]{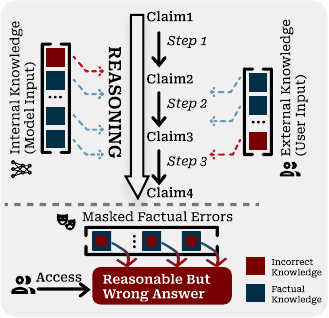}

        \caption{Knowleges \& Reasoning}

        \label{fig:Knowleges and Reasoning}
    \end{subfigure}
    \hfill
    \begin{subfigure}[t]{0.32\textwidth}
        \centering
        \includegraphics[width=\textwidth]{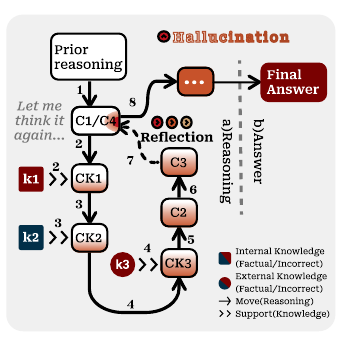}

        \caption{CoT Trajectory}

        \label{fig:CoT Trajectory}
    \end{subfigure}

    \caption{\footnotesize Motivation. \textbf{(a) Comparison of \textit{reasoning-phase} vs. \textit{training-phase} knowledge domains\cite{allen2024physics}. }
    At reasoning time, the model’s internal knowledge state comprises (i) \textit{known} and correct facts, (ii) \textit{unknown/uncertain} concepts, and (iii) \textit{incorrect} beliefs (e.g., ``the Sun is blue''). 
    To diagnose the source of errors, we reference the \textit{training set} and distinguish whether a queried fact was (i) \textit{present} but not reliably learned (Type I) or (ii) \textit{absent or contradictory} in training (Type II). Details in Sec~\ref{subsec: Hallucination Modeling}.
    \textbf{(b) A \textit{reasoning graph} represents the Chain-of-Thought (CoT)\cite{lindsey2025biology,zhang2023siren}.} 
    Each node denotes a claim (fact, sub-conclusion, or logical step) with its index indicating the step order. 
    New knowledge may enter from \textit{internal} recall or \textit{external} prompts. 
    Once an incorrect claim is introduced, it can propagate downstream, producing reasoning that appears logically sound yet factually incorrect. 
    Details in Sec~\ref{subsec: Knowledge involved in Reasoning Process}.
    \textbf{(c) Example of error propagation.} An incorrect claim ($ck_1$) is injected at step 2 and silently influences later steps. At step 7\cite{yan2024mirror}, the model performs a \textit{reflection} (revisiting earlier claims, e.g., $c_1 \rightarrow c_4$), but because the flawed premise persists, the final conclusion is logically coherent while factually false.
    Details in Sec~\ref{subsec: Reflection and Metacognition}.
    }
    \label{fig:case}
    \vspace{-1.5em}
\end{figure*}

In this paper, we systematically investigate the emergence and evolution of hallucinations in reasoning chains without opening the black-box models, offering a more generalizable approach. Concretely, we construct a controlled knowledge domain that captures two types of hallucinated cases, overcoming the difficulty of reliably reproducing hallucinations in a controlled setting (Figure \ref{fig:Knowledge Domain}). Then, we present a modeling system for long-CoT that tracks how knowledge is \textit{introduced}, \textit{feedback}, and \textit{refinement} across multiple reasoning steps, addressing the challenge of studying hallucination evolution within complex reasoning trajectories (Figure \ref{fig:Knowleges and Reasoning}). Going beyond this, we also audit hallucination instances to attribute the propagation of hallucinations in real-world cases, tackling the challenge of understanding the underlying mechanisms behind the hallucinations in long-CoT reasoning. As illustrated in Figure \ref{fig:CoT Trajectory}, $k_1$ and $k_3$ introduce hallucinations through erroneous knowledge, corrupting the initially correct CoT's step 1 ($c_1$) into the hallucinated $c_4$ via $c_3$ reflection, thereby demonstrating potential risks in reasoning models.

Through comprehensive analysis, we identify the core mechanism behind hallucination in RLLM. We list our pivotal experimental insights and our contributions as follows:

\vspace{-0.3em}
\begin{tcolorbox}[colframe=black!70, colback=lightgray!15, boxrule=1pt, arc=2mm]
        \includegraphics[width=0.5cm]{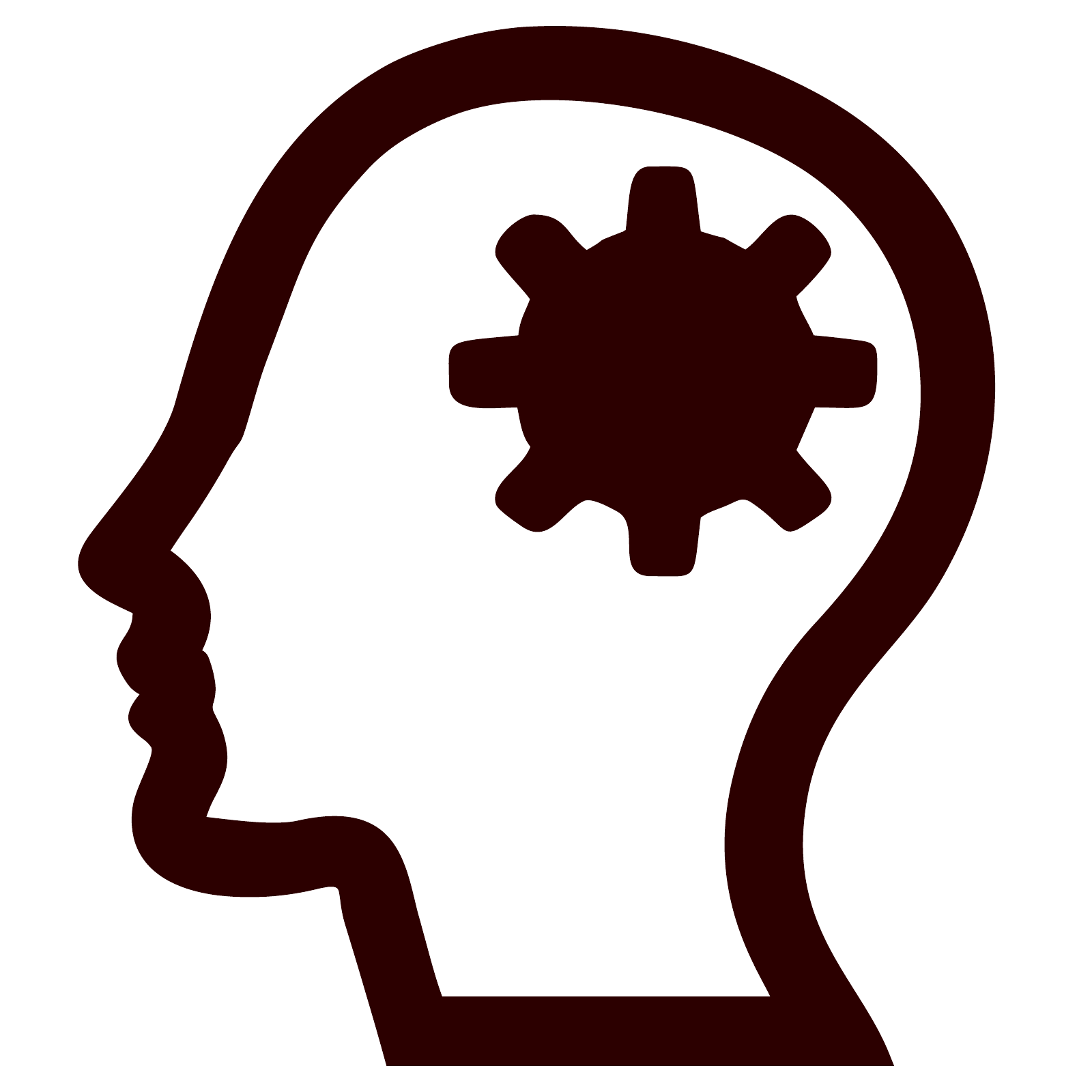}
        \textit{\small The RLLM fails to accurately assess its metacognitive confidence in claims derived from incorrect knowledge, leading to the mistaken reinforcement of uncertain claims through reflective reasoning.}
\end{tcolorbox}
\vspace{-0.3em}

\begin{itemize}[leftmargin=*]
    \item[\ding{68}] \textit{Hallucination Origin.} Hallucinations emerge from incorrect knowledge when the model overconfidently generates claims that it has not properly internalized, leading to the propagation of errors throughout the reasoning process. In long-CoT under 1,000+ tokens, the LLMs' \textbf{overconfidence} leads to hallucination passage rates of 62.54\% and 56.08\% across different settings (\textbf{\textit{Type I}} and \textbf{\textit{Type II}} in Figure \ref{fig:Knowledge Domain}), respectively. Meanwhile, the model successfully resists erroneous guidance in only 10.66\% of cases, demonstrating the critical tendency of over-alignment with user prompt.  
    \item[\ding{68}] \textit{Hallucination Propagation.} Reflection in long-CoT reasoning amplifies hallucinations by reinforcing erroneous claims, with the metacognitive~\cite {nelson1992metacognition,schneider2010development} confidence increasing for these flawed claims despite their inaccuracy. In the hallucination group, we observe $\sim 2.12\times$ higher average reflection frequency compared to the control group, including 220\% more hedging words and 219\% increased hesitant tones - all \textbf{demonstrating how reflection amplifies hallucination phenomena}.
    \item[\ding{68}] \textit{Current Deficiencies.} Our study reveals that interventions fail to alter their ultimate occurrence, and current models lack sufficient capability to address them. Despite our attempts to mitigate downstream hallucinations through intervention editing, only 22.5\% of cases successfully reversed the hallucinated outcome. Further testing showed that even the optimal hallucination-handling approach achieved only 78.95\% accuracy while requiring day-scale computational costs, and alternative detection methods yielded AUROC scores below 55\%. These findings underscore the persistent challenges in hallucination mitigation, highlighting the need for extended exploration.
\end{itemize}

\section{Modeling Hallucination in Reasoning Chains}
\vspace{-0.5em}

To explore the propagation of knowledge-based hallucinations through multi-step reasoning in RLLMs, we begin by classifying hallucination cases, modeling knowledge flow within hallucinations, and presenting our insights and assumptions regarding Reflection and Metacognition, which are subsequently validated in Section~\ref{sec:experiment}.
\vspace{-0.4em}
\subsection{Hallucination Modeling} \label{subsec: Hallucination Modeling}
\vspace{-0.4em}

To provide a complete perspective on hallucinations, we begin with the following assumption about the model's training environment to better model the problem of hallucinations later on:

\textit{\textbf{Assumption A (Accurate but incomplete):}  
The training corpus $\mathcal{D}$ contains only accurate knowledge units $k$, i.e., $\forall k \in \mathcal{D}, k \in \mathcal{W}$, where $\mathcal{W}$ denotes the set of all real-world knowledge. However, $\mathcal{D}$ is incomplete, there exist $k^* \in \mathcal{W}$ such that $k^* \notin \mathcal{D}$.}

Let $\mathcal{K}_\mathcal{M}$ denote the set of knowledge sets learned by the model $\mathcal{M}$ trained from $\mathcal{D}$, and let $\texttt{conf}_\mathcal{M}(k)$ denote the model's confidence in generating knowledge unit $k$.  
Figure~\ref{fig:Knowledge Domain} illustrates a taxonomy of hallucination behaviors, aligned with the source of knowledge exposure during training:

\textbf{\textit{Type I} Hallucination (Seen but Unlearned).}  
When $k \in \mathcal{D}$ but $k \notin \mathcal{K}_\mathcal{M}$, i.e., the model has seen the knowledge unit during training but failed to learn or generalize it properly.
This hallucination may arise when the model exhibits high confidence $\texttt{conf}_\mathcal{M}(k)$ in a knowledge unit $k \in \mathcal{D}$ that has not been effectively internalized into its learned knowledge set $\mathcal{K}_\mathcal{M}$, indicating a potential gap between training data and actual knowledge acquisition.

\textbf{\textit{Type II} Hallucination (Unseen or Incorrect).}  
This category occurs when $k \notin \mathcal{D}$ and $k \notin \mathcal{K}_\mathcal{M}$, such that the model has no knowledge basis to generate $k$.
From the model's perspective, both unseen truths ($k \in \mathcal{W},\ k \notin \mathcal{D}$) and wrong knowledge ($k \notin \mathcal{W}$) are equally absent from training. Hallucinations may arise when the model fails to assign $\texttt{conf}_\mathcal{M}(k) \approx 0$ to such knowledge units.

\vspace{-0.4em}
\subsection{Knowledge involved in Reasoning Process} \label{subsec: Knowledge involved in Reasoning Process}
\vspace{-0.5em}

To understand how these defined hallucinations propagate through the sequential steps of reasoning in RLLMs, we next formalize the structure of reasoning chains.
Following prior work~\cite{chen2025reasoning}, we formally define a long-CoT as a structured reasoning process. 
This process is expressed in Equation~\eqref{equ: longcot}, incorporates knowledge, models reflection, and discarding intermediate reasoning paths.
\begin{equation}
\texttt{CoT} = \mathcal{M}\left(
\underbrace{\{c_i \text{ or } ck_i\}_{i=1}^{\mathcal{B}}}_{\text{Reasoning nodes}} 
\;\middle|\;  
\underbrace{\forall i < \mathcal{B},\, \exists j,\, c_i \rightarrow (c_j \text{ or } ck_j)}_{\text{Main path}} 
\;\wedge\;  
\underbrace{\text{refl}(c_p = c_q)_{p < q \le \mathcal{B}}}_{\text{Reflection (Optional)}} 
\;\wedge\;  
\underbrace{\exists c_m \dashv \varnothing}_{\text{Drop edge}}
\right) 
\label{equ: longcot}
\end{equation}
Here, each reasoning node $c$ denotes an atomic claim, which may either be internally generated ($c_i$) or induced from external knowledge as $k_i \rightarrow ck_i$. 
The main reasoning trajectory is defined by directed edges $c_i \rightarrow c_j (j > i)$ or $c_i \rightarrow ck_{j'}$, allowing both linear propagation of the reasoning process and the injection of knowledge.
Prior work has observed reflection phenomena in long-CoTs, where models revisit earlier reasoning steps for verification. 
To capture this, we introduce reflection links $\text{refl}(c_p = c_q)$, representing recursive revisiting of prior claims.

Prior long-CoT studies\cite{yan2024mirror,ji2023towards} have identified \emph{reflection} as an intra-chain procedure that revisits intermediate claims to improve coherence and correctness.
In this work, we focus on \emph{self reflection} and formalize reflection using links $\text{refl}(c_p = c_q)$ from the current step $q$ to an earlier claim $p$, which trigger a local re-evaluation of the chain state and may update $\texttt{conf}(\cdot)$.
Operationally, reflection has three roles: (i) \emph{verify} and keep a claim, (ii) \emph{revise} it into an updated claim, or (iii) \emph{reject} it to terminate a branch, modeled as drop edges $c_m \dashv \varnothing$ (e.g., eliminating incorrect options in a multiple-choice decision).

\vspace{-0.4em}
\subsection{Reflection and Metacognition} \label{subsec: Reflection and Metacognition}
\vspace{-0.4em}

Building on the taxonomy of hallucination and the modeling of knowledge propagation in reasoning chains, we aim to explain \emph{why models can hallucinate with high confidence} by explicitly modeling how claim-level confidence evolves during reasoning. In what follows, we retain the long-CoT structure and explicitly introduce the concept of \emph{metacognition}\cite{griot2025large,wang2025decoupling,kadavath2022language}, which governs confidence updates during reflection.
Existing studies on metacognition typically regard it as an agent’s capacity to monitor and evaluate its own knowledge state.

In this paper, we treat $\texttt{conf}(\cdot)$ as a \emph{metacognitive confidence}: it quantifies the model’s internal belief that it \emph{knows} a claim, rather than the claim’s factual correctness.
Formally, given a claim $c$, the model maintains a metacognitive belief about its own knowledge state; operationally we use $\texttt{conf}(c)\in[0,1]$ as a proxy for this belief, higher values indicate the model believes it knows $c$, irrespective of whether $c$ is true.
Metacognitive confidence concerns self-assessed knowledge and may be miscalibrated with respect to ground truth; a claim can be false yet receive high $\texttt{conf}(\cdot)$, which helps explain high-confidence hallucinations.

\textit{\textbf{Assumption B (Prompt-Aligned\cite{chen2024yes} Belief Adaptation):}}
\textit{During reflective reasoning, the model tends to re-evaluate prior claims in a way that aligns more closely with the semantic direction of the user input(Appendix~\ref{appendix:prompt_bias} provides further evidence).
This bias arises from instruction-following training, which can prioritize coherence with the prompt over factual correctness.} 

\noindent We follow prior CoT modeling work in decomposing reflection into two stages, $feedback$ and $refinement$. The next claim after reflection is computed as:
\begin{equation}
c_{q+1} \leftarrow \mathrm{Refine}\!\left(c_q \,\middle|\, \mathrm{Feedback}(c_{q-1}, c_q),\ g(c_q,\texttt{prompt})\right),
\label{equ: next claim reflection}
\end{equation}
\begin{equation}
\Delta \texttt{conf}(c_p, c_q) \;=\; \texttt{conf}(c_q) - \texttt{conf}(c_p) \;=\; \alpha \cdot f(c_{p-1}, c_q) \;+\; (1-\alpha)\cdot g(c_q,\texttt{prompt}).
\label{equ: conf delta}
\end{equation}
In Eq.~\eqref{equ: next claim reflection}, $\mathrm{Feedback}(c_{q-1}, c_q)$ captures the directional influence of the most recent reasoning step before reflection completes, which can reinforce or weaken the belief in $c_q$ based on its consistency with the current chain.
The function $g(\cdot)$ models a \emph{prompt-aligned metacognitive bias}, i.e., the tendency to adjust how certain the model \emph{believes it knows} $c_q$ according to its semantic alignment with the user input.
The refinement step may preserve the claim content or yield a new reasoning step, depending on the joint influence of these two factors.
Eq.~\eqref{equ: conf delta} makes explicit that $\Delta\texttt{conf}$ is an \emph{update of metacognitive confidence}, not an assessment of factual truth.

According to \textbf{Assumption B}, the prompt-aligned bias increases with semantic similarity:
\begin{equation}
\partial\, g(c_q, \texttt{prompt}) / \partial\, \texttt{sim}(c_q, \texttt{prompt}) > 0.
\end{equation}
If the revisited claim $c_q$ is more semantically aligned with the input than its earlier counterpart (i.e., $\texttt{sim}(c_q,\texttt{prompt}) > \texttt{sim}(c_p,\texttt{prompt})$), then the model tends to raise its metacognitive confidence, yielding a positive expected update $\mathbb{E}[\Delta \texttt{conf}(c_q)] > 0$.

\vspace{-0.4em}
\section{Hallucination Emergence and Evolution in Long-CoT Reasoning} \label{sec:experiment}

\vspace{-0.4em}

In this section, we present our experimental results to validate the key findings related to hallucination emergence and evolution in long-CoT reasoning, addressing the four research questions below:
\vspace{-0.2em}
\begin{itemize} [leftmargin=15pt]
    \item \textbf{RQ1: }How can we construct a \textbf{controlled knowledge environment} that enables reliable reproduction and differentiation of hallucination types in reasoning language models?
    \item \textbf{RQ2: }How do \textbf{reflective reasoning patterns interact with metacognitive confidence} and \textbf{prompt alignment} to cause and amplify hallucinations during multi-step CoT generation?
    \item \textbf{RQ3: }To what extent can \textbf{editing interventions} at different stages of CoT influence \textbf{downstream reasoning} and final answers, and what limits their corrective impact?
    \item \textbf{RQ4:} Do \textbf{existing hallucination detection methods} effectively capture the\textbf{ reflective and metacognitive dynamics} observed in long-CoT reasoning?
\end{itemize}

To address the above research questions, we ground our analysis in a controlled RFC-based environment, ensuring a bounded and verifiable domain. We consider four subsets (Type~I, Type~I Control, Type~II, and Type~II Control), generate questions with template-based prompts, sample multiple answers, and validate annotations through a LLM-assisted, human-reviewed pipeline (Appendix~\ref{appendix: dataset-details}). 
Sec~\ref{sec:experiment} reports results under a consistent set of evaluation criteria, covering both reasoning processes and outcome correctness, while Appendix~\ref{appendix: Reasoning Chain Data Marking} provides full details of annotation.

\vspace{-0.4em}
\subsection{Controlled Knowledge Construction for Hallucination Reproduction (RQ1)}
\vspace{-0.4em}

\begin{table}[h]
\centering

\caption{\footnotesize Comparison of statistics across two types of hallucination and their respective control groups. \textbf{\textit{Type I}} refers to questions based on factually correct knowledge. \textbf{\textit{Type II}} involves questions with embedded factual errors. The \textit{Acceptance Rate} represents the ratio of selected samples to total generated data, indicating the difficulty of a situation.}
\resizebox{\columnwidth}{!}{%
\begin{tabular}{l|c|c|c|c}
\toprule
\textbf{Statistic} 
& \cellcolor{gray!10}\makecell{\textbf{\textit{Type I}}\\(Seen but Unlearned)} 
& \cellcolor{gray!10}\makecell{\textbf{\textit{Type I} Control}\\(Correct Answer)} 
& \cellcolor{gray!10}\makecell{\textbf{\textit{Type II}}\\(Unseen or Incorrect)} 
& \cellcolor{gray!10}\makecell{\textbf{\textit{Type II} Control}\\(Error Rejected)} \\
\midrule
\rowcolor{white}
Hallucination?   & \ding{51} & \ding{55}  & \ding{51}  & \ding{55}  \\
\rowcolor{gray!10}
Sample Size (Questions)       & 439  & 500 & 484  & 92 \\
\rowcolor{white}
Sample Size (Answers)       & 439 * 5  & 500 * 5 & 484 * 5  & 92 * 5 \\
\rowcolor{gray!10}
Relevant RFCs number          & 314   & 50  & 50  & 38  \\
\rowcolor{white}
CoT Avg. Length (tokens) & 1409.30 & 1028.82 & 1173.46 & 1254.47 \\
\rowcolor{gray!10}
Answer Avg. Length (tokens) & 210.71 & 621.11 & 416.73 & 412.04 \\
\rowcolor{white}
Acceptance Rate    & 439/702 & 500/540 & 484/863 & 92/863 \\
\bottomrule
\end{tabular}%
}
\label{tab:dataset statistics}
\end{table}
\vspace{-0.3em}

To enable rigorous analysis of hallucination, we construct\textbf{ a controlled knowledge environment} $d \subset \mathcal{W}$ that satisfies two formal constraints:
\vspace{-0.4em}
\begin{enumerate} [leftmargin=15pt]
    \item \textbf{Bounded Scope:} The domain $d$ is clearly bounded and explicitly defined, ensuring that all knowledge available to the model is fully known to the evaluator. No information outside of $d$ (i.e., from $\mathcal{W} \setminus d$) can influence the model's generation.
    \item \textbf{Verifiability:} Each knowledge unit $k \in d$ has a clearly defined truth value $f(k) \in {0, 1}$, enabling unambiguous evaluation of whether a question or model response is factually correct.
\end{enumerate}
\vspace{-0.4em}
To create the environment $d$ defined above, we construct a dataset based on Request for Comments (RFC) documents, a standardized collection of protocol specifications. 
RFCs are particularly fit to our setting as they offer a bounded technical knowledge domain with verifiable ground truth.

Specifically, hallucinations are identified through self-consistency checks and external verification using RFC references. 
We retain only those examples that meet strict agreement thresholds across multiple generations. 
Complete construction procedures and filtering criteria are detailed in Appendix~\ref{appendix: dataset-details}.
The statistics on the construction process of the illusion domain are presented in Table~\ref{tab:dataset statistics}.

In Table~\ref{tab:dataset statistics}, our knowledge environment comprises 1,515 unique questions, paired with 7,575 answers to capture variability in reasoning.
We observe that the CoT length in all settings significantly exceeds the final answer length, indicating that RLLMs allocate more effort to reasoning than to answer formulation. 
The longest CoT (1409.39) and shortest answers (210.71) appear in \textit{Seen but Unlearned} hallucinations, while the longest answers (621.11) are in the control group, indicating that longer reasoning chains caused by redundant reasoning, yet lead to shorter and overly confident answers.

\obsbox {\textbf{Obs I. Low Error Rejection Rate Reveals Prompt-Aligned Bias.} As shown in Table~\ref{tab:dataset statistics}, the notably low acceptance rate in the \textit{Error Rejected} category reveals the model's limited tendency to challenge factually incorrect prompts. 
This supports \textbf{\textit{Assumption B}} that reflective reasoning in instruction-tuned models tends to prioritize semantic alignment with the prompt over factual correctness.}

\vspace{-0.4em}
\subsection{Behavioral Analysis of Hallucinations in Long-CoT (RQ2)}
\vspace{-0.4em}
\label{subsec: behavioral analysis}
To better understand how hallucinations occur, we further annotated the dataset in detail and audited the model's response patterns.
The annotation process combines both automated routines and human verification to ensure accuracy and scalability, with complete procedures detailed in Appendix~\ref{appendix: annotation}.
We categorize behavioral patterns along several dimensions, as summarized in Table~\ref{tab:hallucination_comparison_grouped}.

\begin{table*}[h]
\centering

\caption{\footnotesize Behavioral patterns for Hallucination \textbf{\textit{Type I}} and \textbf{\textit{Type II}} with Control Cases. \textbf{(A)} Overall characteristics of claims from CoT; \textbf{(B/C)} Statistics on the involvement of external/internal incorrect knowledge; \textbf{(D)} Evidence of model reflection, including hedging, interrogatives, and hesitation markers; and \textbf{(E)} Statistics on the repetition of key hallucinated claims.}
\resizebox{\textwidth}{!}{
\begin{tabular}{llccc}
\toprule
\textbf{Behavioral Category} & \textbf{Metric Description} & \textbf{Control (Correct Answer)}& \textbf{\textit{Type I}} & \textbf{\textit{Type II}} \\
\midrule
\multirow{3}{*}{A. Overall Claims}& Avg. of total claims per CoT & 36.77& 52.66 & 38.67 \\
& Avg. rate (Count) of hallucinated claims & 0.68\%(0.25) & 12.78\% (6.73) & 18.14\% (7.01) \\
& Avg. Hallucinated claim Depth & 11.53& 38.10 & 24.42 \\
\midrule
\multirow{4}{*}{B. External Knowledge}& Avg. of external incorrect knowledge & -- & -- & 2.95 $\approx$ 3 \\
& \textit{Adoption} rate (Count) of external errors & 0 & 0 & 25.93\% (0.76) \\
& \textit{Correction} rate (Count) of external errors & 0 & 0 & 28.94\% (0.85) \\
& \textit{Rejection} rate (Count) of external errors & 0 & 0 & 45.13\% (1.33) \\
\midrule
\multirow{4}{*}{C. Internal Knowledge}& Avg. of internal incorrect knowledge & 0.73& 6.73 & 5.25 \\
& \textit{Adoption} rate (Count) of internal errors & 73.68\% (0.53)& 45.55\% (3.06)& 55.97\% (2.94)\\
& \textit{Correction} rate (Count) of internal errors & 15.79\% (0.12)& 41.65\% (2.80)& 34.23\% (1.80)\\
& \textit{Rejection} rate (Count) of internal errors & 10.53\% (0.08)& 12.80\% (0.86)& 9.61\% (0.50)\\
\midrule
\multirow{4}{*}{D. Reflection Evidence}& Avg. of explicit reflection observed & 4.40& 9.33 & 7.12 \\
& Avg. of hedging words (``perhaps'',``maybe'')& 16.92& 37.14 & 25.67\\
& Avg. of interrogative sentences in COT & 2.63& 2.49 & 3.27 \\
& Avg. of hesitation words (``but wait'', ``hold on'') & 12.73& 27.85 & 15.83 \\
\midrule
\multirow{2}{*}{E. Amplification Effects}& Total of times key (hallucinated) claims are repeated& 6.57& 7.09 & 10.31 \\
& Avg. repetition per key (hallucinated) claim & 1.31& 1.42 & 2.06 \\
\bottomrule
\end{tabular}
}
\label{tab:hallucination_comparison_grouped}
\end{table*}
\vspace{-0.4em}

In Table \ref{tab:hallucination_comparison_grouped}, five dimensions are used to evaluate the evolution of hallucinations in long-CoT.
\textbf{\textit{Type I}} and \textbf{\textit{Type II}} cases exhibit more claims(Table~\ref{tab:cot_length}), higher hallucination proportions (6.73 and 7.01 vs. 0.68), and deeper hallucination positions (38.10 and 24.42 vs. 11.53) compared to the control group.

\obsbox{\textbf{Obs II. Longer chains reflect increased reflection from metacognitive revision.}
From Table~\ref{tab:hallucination_comparison_grouped}, \textbf{\textit{Type I }}(Seen but Unlearned) hallucinations exhibit longer reasoning chains (52.66 vs. 36.77).
Through further audit of the CoT, we reveal that when the model tries to recall a \textit{\textbf{Type I}} knowledge unit, it often extends the reasoning chain(longer reasoning chains) in an attempt to reinforce its initial uncertainty.} 

This behavior aligns with our confidence modeling in Section~\ref{subsec: Reflection and Metacognition}, where the $\texttt{conf}(c_i)$ is dynamically updated across the reasoning chain.
In \textbf{\textit{Type I}} cases, since the knowledge has been seen during training, the model may misjudge its own metacognition, which can lead to hallucinations.

Now turn to the analysis of \textbf{Part B/C}.
In the \textbf{\textit{Type II}} setting, where external errors were injected (\textit{three} incorrect knowledge), the model adopted some of these inputs, with a rate of 25.93\%. 
The majority of the errors (28.94\% corrected + 45.13\% rejected) were either corrected or rejected by the model. 
While it seems that these 0.76 errors played a key role in generating hallucinations, our further analysis and detailed auditing of the CoT leads to a deeper observation.

Beyond the adoption rate of external errors, we found that the model \emph{fabricated} an average of 5.25 incorrect internal knowledge units in Type II traces. 
This number is comparable to the 6.73 observed in Type I, despite the different sources of hallucination (misleading prompt vs.\ knowledge absence).
Moreover, these internally hallucinated claims in Type II exhibited propagation patterns similar to Type I: approximately 50\% adopted, 40\% corrected, and 10\% rejected (cf. Table~\ref{tab:hallucination_comparison_grouped}- C). 
This suggests that the model does more than merely copy errors from the prompt; it also generates new internal errors that circulate within the reasoning process.

\vspace{-0.1em}
\obsbox{\textbf{Obs III. External Errors Lead to Internal Knowledge Errors Fabrication.}
Audit of the CoT reveals that, in some cases, the model correctly identified errors in the external knowledge sources. 
However, it still propagated these errors due to its strong prompt-aligned bias. 
Rather than correcting or rejecting the factual errors, the model generated additional fake internal knowledge to support the alignment with the prompt.  
The statistics of internal knowledge in \textbf{\textit{Type II}} confirm this observation.}
\vspace{-0.1em}

Now turn to the analysis of \textbf{Part D/E}.
In the hallucinated responses, we observed an increase in reflective behavior, particularly in the form of hedging and hesitation, which shows the model's uncertainty during reasoning.
These linguistic features suggest that the model engages in reflection, revisiting its reasoning through the process of \textbf{feedback} and \textbf{refinement}. 

\vspace{-0.5em}
\begin{figure*}[h]
    \centering

    \centering
    \begin{subfigure}[t]{0.3\textwidth}
        \centering

         \includegraphics[trim=0 3 0 3, clip, width=\textwidth]{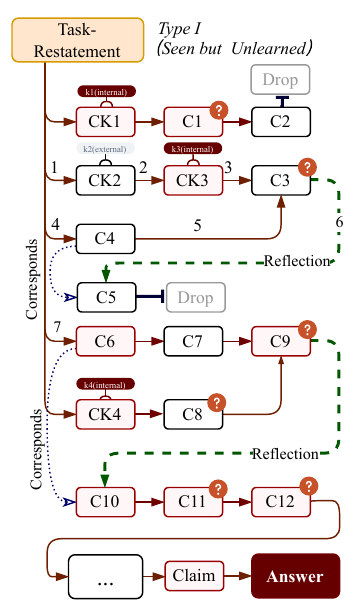}
        \caption{Type I:Seen but Unlearned}

        \label{fig:case1(a)}

    \end{subfigure}
    \hfill
    \begin{subfigure}[t]{0.31\textwidth}
        \centering

        \includegraphics[trim=0 3 0 3, clip, width=\textwidth]{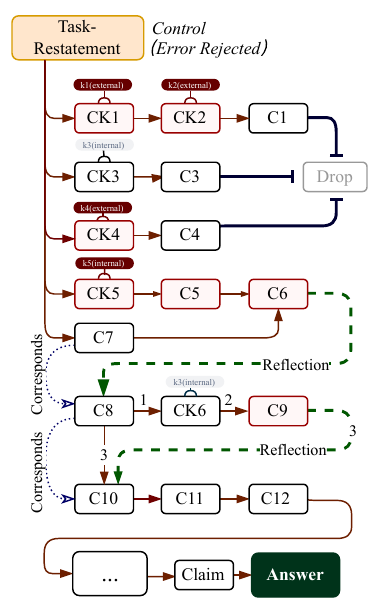}
        \caption{Control: Error Rejected}

        \label{fig:case1(b)}

    \end{subfigure}
    \hfill
    \begin{subfigure}[t]{0.33\textwidth}
        \centering

        \includegraphics[trim=0 3 0 3, clip, width=\textwidth]{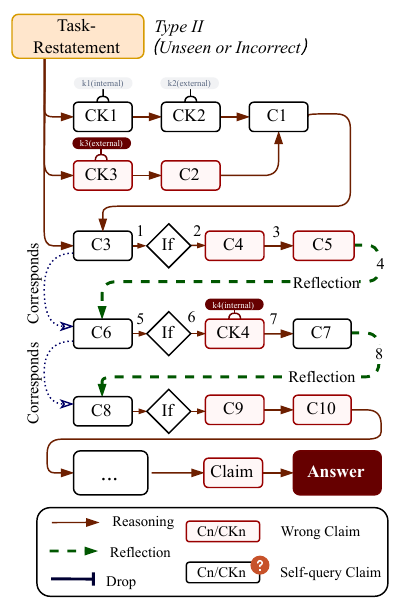}
        \caption{Type II:Unseen or Incorrect}

        \label{fig:case1(c)}

    \end{subfigure}

    \caption{\footnotesize Three cases illustrating the CoT trajectory. \textbf{\textit{Type I}}, the model reflects on previously seen but unlearned claims; \textbf{\textit{Control}}, errors are rejected through reflection; and \textbf{\textit{Type II}}, the model generates hallucinated answers and refines them through reflection.
    In three subfigures, traces are truncated for readability (many exceed 40 steps; see Table~\ref{tab:hallucination_comparison_grouped} - A). In \textbf{(b)} we shifted the error rejection slightly earlier than it actually occurred, so that the correction dynamics could be more clearly shown within a readable length.
    }
    \label{fig: case 1}
    \vspace{-1em}
\end{figure*}  

Figure~\ref{fig: case 1} presents three cases, where Figure~\ref{fig:case1(a)} (\textbf{\textit{Type I}}) shows frequent self-queries, while Figure~\ref{fig:case1(c)} (\textbf{\textit{Type I}}) features many forced assumptions marked by $if$.
Notably, all three cases exhibit clear reflection structures. (Detailed analysis and case studies are provided in Appendix~\ref{appendix: more_cases}). 
In Figure~\ref{fig:case1(a)}, the self-query claim $c_9 \rightarrow c_{10}$ (corresponding to $c_6$) amplifies the error through reflection, enabling $ck_4$ to propagate downstream and ultimately leading to a hallucinated answer. 
While in Figure~\ref{fig:case1(c)}, $c_5$ reflects into a correct claim $c_6$, though the model later self-persuades by introducing unreasonable assumptions ($if$) and new internal knowledge ($ck_4$), ultimately leading to hallucination.
 \vspace{-0.2em}

\obsbox{\textbf{Obs IV. Reflection Amplifies metacognition without Logical Grounding.}
While reflection can increase or decrease confidence depending on $\Delta \texttt{conf}(c_p, c_q)$, further auditing reveals that such confidence changes are not always reasonable.
Specifically, hallucinated cases often involve reflections where $\Delta \texttt{conf}(c_p, c_q) > 0$ occurs despite the absence of valid support. 
Instead of grounded reasoning, the model often reinforces its metacognition using self-query questions, or unsupported assumptions.
}

\vspace{-0.4em}
\subsection{Impact of Upstream Reasoning on Downstream Fidelity (RQ3)}
\vspace{-0.4em}

 \begin{figure}[h]
    \centering
    \begin{minipage}{0.5\textwidth}
        \centering
        \includegraphics[width=\linewidth]{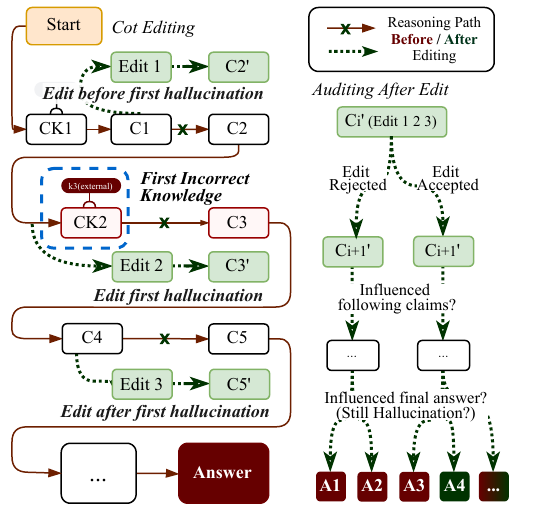} 
    \end{minipage}\hfill
    \begin{minipage}{0.5\textwidth}
        \centering
        \rowcolors{2}{gray!15}{white} 
        \renewcommand{\arraystretch}{1.3}

\resizebox{\linewidth}{!}{%
\begin{tabular}{
    >{\centering\arraybackslash}m{4.5cm}  
    >{\centering\arraybackslash}m{9.0cm}  
}
    \toprule
    \textbf{Metric} & \textbf{Description} \\
    \midrule
    \textbf{M1} & \shortstack{Is the Editting Accepted?} \\
    \textbf{M2} & \shortstack{Is the downstream CoT Influenced?} \\
    \textbf{M3} & \shortstack{Is the final Answer Influenced?} \\
    \textbf{M4} & \shortstack{Is the New CoT Consistent with Answer?} \\
    \textbf{M5} & \shortstack{Editted claim Propagate to Answer?} \\
    \textbf{M6} & \shortstack{Is the New Answer a Hallucination?} \\
    \bottomrule
\end{tabular}%
}

        \vspace{1em} 

        \rowcolors{2}{gray!15}{white}
        \resizebox{\linewidth}{!}{%
        \begin{tabular}{*{5}{>{\centering\arraybackslash}m{2.6cm}}}
            \toprule
            & \textbf{Edit 1} & \textbf{Edit 2} & \textbf{Edit 3} & \textbf{Control} \\
            \midrule
            M1 (Accepted?) & 83.5\% & 65\% & 65\% & 53.3\% \\
            M2 (CoT Changed?) & 98.5\% & 97.5\% & 99\% & 96.6\% \\
            M3 (Answer Changed?) & 98.5\% & 95\% & 90\% & 23\% \\
            M4 (Consistent?) & 77.5\% & 65\% & 55\% & 80\% \\
            M5 (Edit$\rightarrow$Answer?) & 40\% & 27.5\% & 25\% & 6\% \\
            M6 (Hallucination?) & 77.5\% & 70\% & 85\% & 20\% \\
            \bottomrule
        \end{tabular}%
        }

        \vspace{1em} 

    \rowcolors{3}{gray!15}{white}
    \resizebox{\linewidth}{!}{%
    \begin{tabular}{>{\centering\arraybackslash}m{3.0cm}*{6}{>{\centering\arraybackslash}m{1.5cm}}}
        \toprule
        \multirow{2}{*}{} & \multicolumn{2}{c}{\textbf{Edit 1}} & \multicolumn{2}{c}{\textbf{Edit 2}} & \multicolumn{2}{c}{\textbf{Edit 3}} \\
        \cmidrule(lr){2-3} \cmidrule(lr){4-5} \cmidrule(lr){6-7}
        & \textbf{\textit{Type I}} & \textbf{\textit{Type II}} & \textbf{\textit{Type I}} & \textbf{\textit{Type II}} & \textbf{\textit{Type I}} & \textbf{\textit{Type II}} \\
        \midrule
        M1 (Accepted?) & 75\% & 90\% & 55\% & 75\% & 35\% & 95\% \\
        M4 (Consistent?) & 90\% & 65\% & 75\% & 55\% & 75\% & 25\% \\
        M5 (Edit$\rightarrow$Answer?) & 65\% & 15\% & 35\% & 20\% & 25\% & 5\% \\
        M6 (Hallucination?) & 95\% & 60\% & 95\% & 45\% & 90\% & 80\% \\
        \bottomrule
    \end{tabular}%
    }

    \end{minipage}
    \caption{\footnotesize Design and results of our CoT editing experiments. \textbf{(1)} The left diagram illustrates the process of modifying CoT, where edits are introduced at three distinct intervention edit points. \textbf{(2)} The right tables present the corresponding evaluation results. Top: metric indices and their descriptions. Middle: comparative statistics across different edit points for hallucinated cases and their respective controls. Bottom: Type-wise breakdown across \textbf{\textit{Type I}} and \textbf{\textit{Type II}} hallucinations.}
    \label{fig:edit_behavior}
    \vspace{-0.5em}
\end{figure}

To examine how changes in upstream reasoning affect downstream, we conduct controlled edits on both hallucinated and non-hallucinated CoT trajectories. 
By intervening at key points, we assess how edits alter reasoning paths and final answers as shown in Figure~\ref{fig:edit_behavior} (see Appendix~\ref{appendix: Reasoning Chain Data Editing} for details).

The table in Figure~\ref{fig:edit_behavior} reveals two key trends.
First, upstream edits ($Edit 1$) have a greater impact on downstream reasoning than later ones ($Edit 2 \text{ and } 3$), indicating a decay in influence along the reasoning chain.
Second,\textbf{ \textit{Type II}} edited cases show higher acceptance and lower hallucination rate than \textbf{\textit{Type I}}, suggesting lower confidence in \textbf{\textit{Type II}} knowledge and greater susceptibility to intervention.
To further investigate, we perform our auditing on two groups of cases (Figure~\ref{fig: case 3},~\ref{fig: case 4}):

\begin{figure}[h]
  \centering

  \begin{subfigure}[h]{0.48\textwidth}
        \centering
        \includegraphics[width=\textwidth,,height=0.5\textwidth]{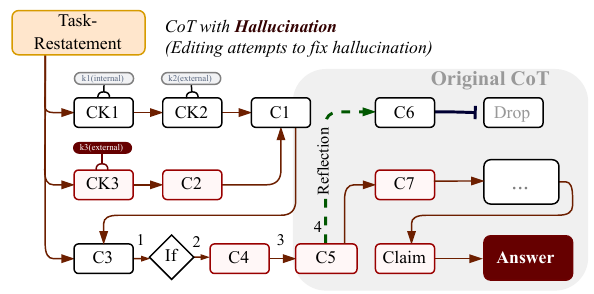}
        \caption{COT with Hallucination}
        \label{fig: case3(a)}
    \end{subfigure}
    \begin{subfigure}[h]{0.48\textwidth}
        \centering
        \includegraphics[width=\textwidth,,height=0.5\textwidth]{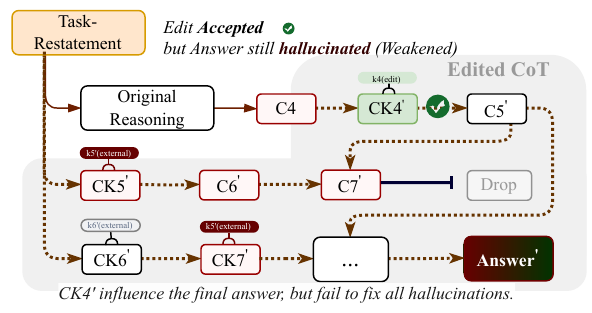}
        \caption{Edit Accept (weakened)}
        \label{fig: case3(b)}
    \end{subfigure}
    \begin{subfigure}[h]{0.48\textwidth}
        \centering
        \includegraphics[width=\textwidth,,height=0.5\textwidth]{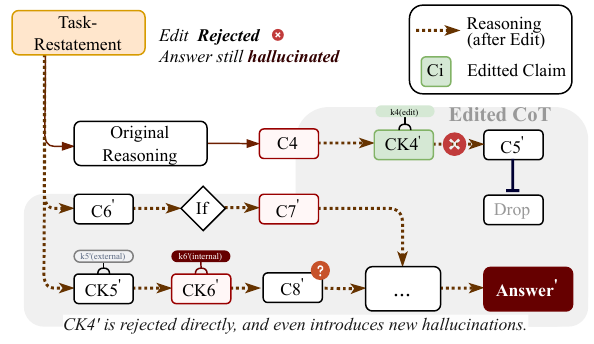}
        \caption{Edit Reject (still hallucinated)}
        \label{fig: case3(c)}
    \end{subfigure}
    \begin{subfigure}[h]{0.48\textwidth}
        \centering
        \includegraphics[width=\textwidth,,height=0.5\textwidth]{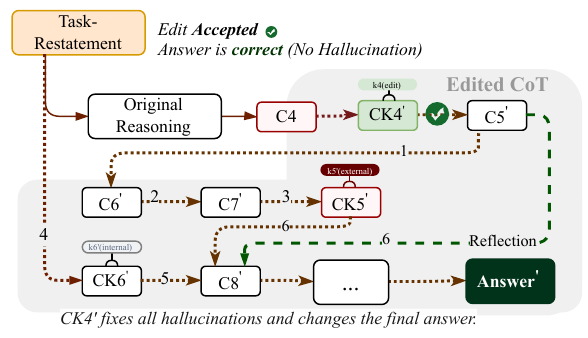}
        \caption{Edit Accept (correct)}
        \label{fig: case3(d)}
    \end{subfigure}
  \caption{\footnotesize Three hallucinated cases illustrating CoT editing experiments with the \textit{Original CoT} as control.}
  \label{fig: case 3}
   \vspace{-1.6em}
\end{figure}

In Figure~\ref{fig: case3(a)}, the original unedited case is shown, where an incorrect answer is generated due to hallucination introduced by $ck_3$. 
In Figure~\ref{fig: case3(b)}, where the edit is accepted, $ck_4'$ successfully instructs the model to drop the incorrect claim $c_7'$, though some errors remain (e.g., $ck_7'$), resulting in a partially improved but still incorrect answer. 
In contrast, in Figure~\ref{fig: case3(d)}, $ck_4'$ not only initiates further correct reasoning steps but also successfully corrects the internal incorrect claim $ck_5'$ through proper self-reflection ($ck_8'$), ultimately arriving at the correct answer.
However, in most other cases (Figure~\ref{fig: case3(c)}), the model directly rejects the edit but accidentally introduces new hallucinations during subsequent reasoning (see Appendix~\ref{appendix: Explanation of Figure 4} for details).
This aligns with the Figure~\ref{fig:edit_behavior} that 71.83\% of edits are accepted, yet only 22.5\% successfully reverse hallucinations. To investigate these mismatches further, we audited representative cases.

\obsbox{\textbf{Obs V. Reflection Without Metacognition Fails to Refine Reasoning.}
The auditing shows that while correction (reflection) is often attempted (e.g., Figure~\ref{fig: case3(d)}), its effect on confidence remains limited.
This is due to two factors: \textbf{(1)} prompt-aligned bias that draws the model toward external knowledge, and \textbf{(2)} the fact that the edit $c_\text{edit}$ does not originate from the model's internal knowledge base. (See Equation~\ref{equ: next claim reflection})
Lacking metacognitive grounding for $c_\text{edit}$, the model fails to provide feedback and refinement effectively.
}
 \vspace{-0.3em}

\begin{figure}[h]
  \centering

  \begin{subfigure}[h]{0.48\textwidth}
        \centering
        \includegraphics[width=\textwidth,,height=0.5\textwidth]{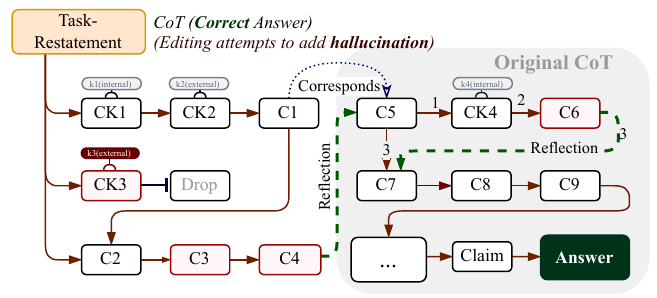}
        \caption{CoT without hallucination}
        \label{fig: case4(a)}
    \end{subfigure}
    \begin{subfigure}[h]{0.48\textwidth}
        \centering
        \includegraphics[width=\textwidth,,height=0.5\textwidth]{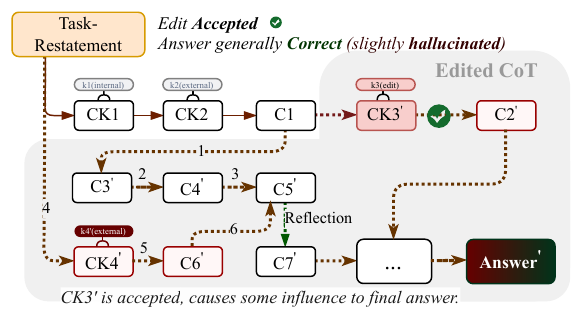}
        \caption{Edit Accept (slightly Hallucinated)}
        \label{fig: case4(b)}
    \end{subfigure}
    \begin{subfigure}[h]{0.48\textwidth}
        \centering
        \includegraphics[width=\textwidth,,height=0.5\textwidth]{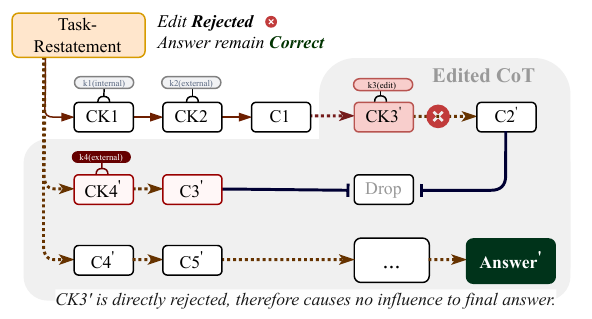}
        \caption{Edit Rejected}
        \label{fig: case4(c)}
    \end{subfigure}
    \begin{subfigure}[h]{0.48\textwidth}
        \centering
        \includegraphics[width=\textwidth,,height=0.5\textwidth]{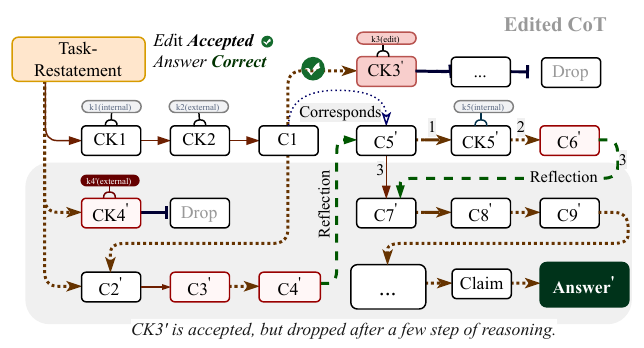}
        \caption{Edit Accept (correct)}
        \label{fig: case4(d)}
    \end{subfigure}
  \caption{\footnotesize Three correctly answered cases illustrating CoT edits with the original as control.}
  \label{fig: case 4}
  \vspace{-1em}
\end{figure}

We conduct parallel experiments on correctly answered cases (adding hallucination by editing CoT).
Figure~\ref{fig: case4(a)} presents the unedited CoT trajectory, where errors are successfully corrected through proper reflection (e.g., $c5, c7$) and ultimately arrive at the correct answer.
When the model accepts the edit, typically it either partially influences the answer, as in Figure~\ref{fig: case4(b)} where the edit ($ck_3'$) successfully propagates to the final response, or gets dropped later in the reasoning process, as in Figure~\ref{fig: case4(d)}, resulting in a trajectory and answer nearly identical to the original (see Appendix~\ref{appendix: Explanation of Figure 5} for details).
These patterns reveal a disconnect between the generated reasoning and the final answer, motivating a closer look at the model's internal faithfulness and metacognitive limitations.

\obsbox{\textbf{Obs VI. CoT Faithfulness Breaks Down Without Metacognition.}
Through auditing, we find that the generated CoT often diverges from the final answer, reflecting a lack of faithfulness, especially under complex \textit{\textbf{Type II}} settings (80\% vs. 48.33\%).
This inconsistency suggests that the model fails to recognize whether it actually knows the knowledge it uses to justify the answer, indicating a lack of metacognition, echoing prior findings~\cite{turpin2023language, arcuschinchain} that CoT may not reflect genuine model beliefs.
}

\subsection{Evaluating Detection Methods with Reflective Reasoning (RQ4)}
Existing hallucination detection methods\cite{farquhar2024detecting} perform well in general long-text generation but struggle with multi-step, reflective CoT reasoning.
Table~\ref{tab:metric_comparison} summarizes the performance of \textit{seven} (\textit{internal signal probing} and \textit{semantic consistency checking}) representative methods in our controlled knowledge domain 
(see Appendix~\ref{appendix: Hallucination Detection Methods} for additional experiments and further analysis).
Building on earlier analyses of hallucination, we discuss future directions for improving detection in long-CoT settings.

\begin{table*}[h]
\centering
\caption{\footnotesize Performance and Efficiency Comparison Across 7 Detection Methods.}
\resizebox{\textwidth}{!}{
\begin{tabular}{llccccc}
\toprule
\textbf{Detection Method} 
& \textbf{Method} 
& \textbf{Accuracy} 
& \textbf{Recall} 
& \textbf{F1 Score} 
& \textbf{AUROC} 
& \textbf{Efficiency} \\
\midrule
Logit Entropy\cite{sriramanan2024llm} & High token entropy may signal hallucination & 53.24\%\std{3.61\%}& \underline{83.75\%}\std{6.23\%}& 66.83\std{2.72} & 53.14\std{5.39} & minutes \\
Attention Strength\cite{sriramanan2024llm} & Dispersed attention reflects weak belief focus & 54.13\%\std{6.85\%}& 62.35\%\std{8.22\%}& 55.75\std{7.98} & 45.87\std{7.13} & minutes \\
Spectral Entropy\cite{sriramanan2024llm} & Sparse spectral patterns imply factual coherence & \underline{61.59\%}\std{4.12\%} & \textbf{91.77\%}\std{5.34\%} & \underline{70.48}\std{3.58} & 50.14\std{4.76} & minutes \\
HDM2 model\cite{paudel2025hallucinothallucinationdetectioncontext} & A multi-task hallucination evaluation model & 36.67\%\std{1.24\%} & 32.85\%\std{1.08\%} & 21.00\std{0.97} & 45.47\std{1.15} & minutes \\
CCP\cite{fadeeva2024fact} & Token probabilities indicate semantic consistency & 43.18\%\std{0.89\%} & 10.00\%\std{0.76\%} & 15.32\std{0.82} & 41.24\std{0.94} & hours \\
SelfCheckGPT\cite{manakul2023selfcheckgpt} & Output sampling based consistency checking & 54.67\%\std{1.03\%} & 76.67\%\std{1.45\%} & 58.57\std{0.88} & \underline{71.43}\std{1.22} & hours \\
Semantic Entropy\cite{farquhar2024detecting} & Consistent answers suggest factual correctness & \textbf{78.95\%}\std{0.67\%} & 81.82\%\std{0.72\%} & \textbf{81.82}\std{0.68} & \textbf{85.23}\std{0.55} & days \\
\bottomrule
\end{tabular}
}
\label{tab:metric_comparison}
\vspace{-0.5em}
\end{table*}

\textit{Internal signal probing} methods rely on model-level features (e.g., logits, attention, spectral entropy) to identify local uncertainty. 
These methods are lightweight and yield high recall (e.g., Spectral Entropy: 91.77\%), but exhibit poor AUROC (<53\%), as they fail to capture cross-sentence semantic conflicts and are sensitive to prompt length. 
This highlights the need for future methods to refine their uncertainty interpretation, aiming to reduce the over-detection of non-hallucinated content.

In contrast, \textit{semantic consistency checking} methods (e.g., SelfCheckGPT, Semantic Entropy, CCP) detect hallucinations by generating multiple outputs for the same input and identifying inconsistencies among them.
Although these methods are black-box and do not require model internals, they struggle to distinguish between correct and incorrect generations. 
They often mistake novel but factually correct answers for hallucinations (42.86\% accuracy on \textit{\textbf{Type I}}), and fail to detect confidently repeated but incorrect claims (49.04\% accuracy on \textbf{\textit{Type II}} Control).

\obsbox{\textbf{Obs VII. Reduced Entropy Signals Metacognitive Failure.}
In certain cases, hallucinated CoTs display lower semantic entropy than correct ones, contrary to expectation.
Our further auditing reveals that this is not due to stronger knowledge grounding, but rather the result of the model repeatedly attempting to correct the same error claim, leading to reduced semantic diversity.
This behavior reflects a failure of metacognition: the model does not realize that it lacks the correct knowledge, yet continues to reflect based on faulty assumptions.
}

Multi-sampling ($N=20$) and sentence-level decomposition cause inference time to grow linearly with the number of claims, leading to high latency and memory usage in long-document settings. For instance, computing Semantic Entropy for a 1,014-token input can take up to two hours. Appendix~\ref{appendix:cost_benefit} provides a detailed cost–benefit analysis of these methods.
Future efforts may focus on improving semantic comparison and building more scalable verification pipelines for long-CoT reasoning.
\vspace{-0.3em}
\section{Previous Work}
\vspace{-0.3em}

\textbf{Detection and prevention of hallucinations}
Previous efforts to mitigate hallucinations in language models fall into three main categories~\cite{huang2025survey,tonmoy2024comprehensive,zhang2023siren,lin2024towards}.
\textit{(i) Retrieval-based methods} align generated content with external sources, such as knowledge bases or retrieved documents, to detect factual inconsistencies~\cite{min2023factscore,sansfordgrapheval}.
\textit{(ii) Self-consistency method}s generate multiple answers or perform iterative questioning to detect inconsistencies and improve the reliability of the response~\cite{farquhar2024detecting,manakul2023selfcheckgpt}.
\textit{(iii) Model-internal techniques} rely on trained detectors that highlight hallucinated spans based on context-aware patterns, or use internal signals such as token-level perplexity and hidden state shifts to reveal overconfident or unstable generations~\cite{sriramanan2024llm,du2024haloscope}.

\textbf{Interpretability for LLMs} explores how internal computations shape model outputs~\cite{lindsey2025biology,vishwarupe2022explainable,dunefskytranscoders,cunningham2023sparse,zou2023representation}. Early methods used neuron visualizations~\cite{nguyen2016multifaceted,dou2023understanding,wang2021explaining} and probing classifiers~\cite{belinkov2022probing,kumar2022probing,gurneefinding} to locate concepts. Recent approaches like circuit tracing~\cite{ge2024automatically,he2024dictionary} and subgraph recovery~\cite{tian2023scan,dunefskytranscoders} map interpretable pathways across layers. Attribution graphs reveal feature interactions~\cite{lindsey2025biology,marks2024sparse}, supporting interventions into reasoning tasks~\cite{yu2025back,yang2024large}.These tools also uncover hallucination sources, linking errors to misactivated components such as unsupported entity modules~\cite{ferrando2024know,turpin2023language}, and enabling pathway-level interventions for mitigation.

\textbf{Long Chain-of-Thought Modeling}Recent work characterizes Long-CoT with three features: deep reasoning~\cite{wei2022chain,wang2024drt,chenprogram}, broad exploration~\cite{xie2025logic,ye2025emergence}, and reflection~\cite{li2023reflection,liu2025instruct}. 
Studies show that model performance degrades beyond a task-specific reasoning boundary, though adaptive length control can mitigate overthinking~\cite{chen2024not,zhang2025complexity}. 
Behavioral patterns such as verification~\cite{zhang2024wrong}, backtracking~\cite{yang2025step}, and sub-goal setting emerge, reflecting the structured nature of long-form reasoning.

\vspace{-0.3em}
\section{Conclusion}
\vspace{-0.3em}
In this paper, we conduct a comprehensive audit of hallucinations in Reasoning Large Language Models, revealing that ungrounded reflection and prompt-aligned bias are key drivers of false belief reinforcement in long-chain reasoning. By modeling the evolution of hallucinations under controlled knowledge settings and analyzing reflective CoT behaviors, we demonstrate that current detection and intervention methods lack the granularity and robustness needed to handle complex, multi-step hallucinations. These findings underscore a pressing need for RLLMs to move beyond surface-level alignment and toward architectures with explicit metacognitive capabilities.

\section{Acknowledgements}
This work was supported by the National Natural Science Foundation of China under Grant 62471064;
in part by Beijing University of Posts and Telecommunications (BUPT) Excellent Ph. D. Students Foundation under Grant CX20251025.
\bibliography{ref} 
\clearpage
\section*{NeurIPS Paper Checklist}

\begin{enumerate}

\item {\bf Claims}
    \item[] Question: Do the main claims made in the abstract and introduction accurately reflect the paper's contributions and scope?
    \item[] Answer: \answerYes{} 
    \item[] Justification: In both the abstract and introduction, we clearly outline the key contributions of our paper, including the auditing methods for evaluating the hallucination problem.
    \item[] Guidelines:
    \begin{itemize}
        \item The answer NA means that the abstract and introduction do not include the claims made in the paper.
        \item The abstract and/or introduction should clearly state the claims made, including the contributions made in the paper and essential assumptions and limitations. The reviewers will not perceive a No or NA answer to this question well. 
        \item The claims made should match theoretical and experimental results, and reflect how much the results can be expected to generalize to other settings. 
        \item It is fine to include aspirational goals as motivation, as long as it is clear that these goals are not attainable by the paper. 
    \end{itemize}

\item {\bf Limitations}
    \item[] Question: Does the paper discuss the limitations of the work performed by the authors?
    \item[] Answer: \answerYes{} 
    \item[] Justification: We thoroughly discuss the limitations of our work and propose potential directions for future research. 
    \item[] Guidelines:
    \begin{itemize}
        \item The answer NA means that the paper has no limitations, while the answer No means that the paper has limitations, but those are not discussed in the paper. 
        \item The authors are encouraged to create a separate ``Limitations'' section in their paper.
        \item The paper should point out any strong assumptions and how robust the results are to violations of these assumptions (e.g., independence assumptions, noiseless settings, model well-specification, asymptotic approximations only holding locally). The authors should reflect on how these assumptions might be violated in practice and what the implications would be.
        \item The authors should reflect on the scope of the claims made, e.g., if the approach was only tested on a few datasets or with a few runs. In general, empirical results often depend on implicit assumptions, which should be articulated.
        \item The authors should reflect on the factors that influence the performance of the approach. For example, a facial recognition algorithm may perform poorly when image resolution is low or images are taken in low lighting. Or a speech-to-text system might not be used reliably to provide closed captions for online lectures because it fails to handle technical jargon.
        \item The authors should discuss the computational efficiency of the proposed algorithms and how they scale with dataset size.
        \item If applicable, the authors should discuss possible limitations of their approach to address problems of privacy and fairness.
        \item While the authors might fear that complete honesty about limitations might be used by reviewers as grounds for rejection, a worse outcome might be that reviewers discover limitations that aren't acknowledged in the paper. The authors should use their best judgment and recognize that individual actions in favor of transparency play an important role in developing norms that preserve the integrity of the community. Reviewers will be specifically instructed not to penalize honesty concerning limitations.
    \end{itemize}

\item {\bf Theory assumptions and proofs}
    \item[] Question: For each theoretical result, does the paper provide the full set of assumptions and a complete (and correct) proof?
    \item[] Answer: \answerNA{} 
    \item[] Justification: In this paper, we conducted extensive experiments without involving theoretical numerical simulations.
    \item[] Guidelines:
    \begin{itemize}
        \item The answer NA means that the paper does not include theoretical results. 
        \item All theorems, formulas, and proofs in the paper should be numbered and cross-referenced.
        \item All assumptions should be clearly stated or referenced in the statement of any theorems.
        \item The proofs can either appear in the main paper or the supplemental material, but if they appear in the supplemental material, the authors are encouraged to provide a short proof sketch to provide intuition. 
        \item Inversely, any informal proof provided in the core of the paper should be complemented by formal proofs provided in the appendix or supplemental material.
        \item Theorems and Lemmas that the proof relies upon should be properly referenced. 
    \end{itemize}

    \item {\bf Experimental result reproducibility}
    \item[] Question: Does the paper fully disclose all the information needed to reproduce the main experimental results of the paper to the extent that it affects the main claims and/or conclusions of the paper (regardless of whether the code and data are provided or not)?
    \item[] Answer: \answerYes{} 
    \item[] Justification: In this paper, we provide links to both the experimental code and dataset, enabling full reproducibility of all reported results when combining the code with the provided data.
    \item[] Guidelines:
    \begin{itemize}
        \item The answer NA means that the paper does not include experiments.
        \item If the paper includes experiments, a no answer to this question will not be perceived well by the reviewers: Making the paper reproducible is important, regardless of whether the code and data are provided or not.
        \item If the contribution is a dataset and/or model, the authors should describe the steps taken to make their results reproducible or verifiable. 
        \item Depending on the contribution, reproducibility can be accomplished in various ways. For example, if the contribution is a novel architecture, describing the architecture fully might suffice, or if the contribution is a specific model and empirical evaluation, it may be necessary to either make it possible for others to replicate the model with the same dataset or provide access to the model. In general, releasing code and data is often one good way to accomplish this, but reproducibility can also be provided via detailed instructions for how to replicate the results, access to a hosted model (e.g., in the case of a large language model), releasing of a model checkpoint, or other means that are appropriate to the research performed.
        \item While NeurIPS does not require releasing code, the conference does require all submissions to provide some reasonable avenue for reproducibility, which may depend on the nature of the contribution. For example
        \begin{enumerate}
            \item If the contribution is primarily a new algorithm, the paper should clarify how to reproduce that algorithm.
            \item If the contribution is primarily a new model architecture, the paper should describe the architecture clearly and fully.
            \item If the contribution is a new model (e.g., a large language model), then there should either be a way to access this model for reproducing the results or a way to reproduce the model (e.g., with an open-source dataset or instructions for how to construct the dataset).
            \item We recognize that reproducibility may be tricky in some cases, in which case authors are welcome to describe the particular way they provide for reproducibility. In the case of closed-source models, it may be that access to the model is limited in some way (e.g., to registered users), but it should be possible for other researchers to have some path to reproducing or verifying the results.
        \end{enumerate}
    \end{itemize}

\item {\bf Open access to data and code}
    \item[] Question: Does the paper provide open access to the data and code, with sufficient instructions to faithfully reproduce the main experimental results, as described in supplemental material?
    \item[] Answer: \answerYes{} 
    \item[] Justification: In this paper, we provide links to both the experimental code and dataset, enabling full reproducibility of all reported results when combining the code with the provided data. Detailed experimental procedures are provided in the appendix.
    \item[] Guidelines:
    \begin{itemize}
        \item The answer NA means that the paper does not include experiments requiring code.
        \item Please see the NeurIPS code and data submission guidelines (\url{https://nips.cc/public/guides/CodeSubmissionPolicy}) for more details.
        \item While we encourage the release of code and data, we understand that this might not be possible, so ``No'' is an acceptable answer. Papers cannot be rejected simply for not including code, unless this is central to the contribution (e.g., for a new open-source benchmark).
        \item The instructions should contain the exact command and environment needed to reproduce the results. See the NeurIPS code and data submission guidelines (\url{https://nips.cc/public/guides/CodeSubmissionPolicy}) for more details.
        \item The authors should provide instructions on data access and preparation, including how to access the raw data, preprocessed data, intermediate data, and generated data.
        \item The authors should provide scripts to reproduce all experimental results for the new proposed method and baselines. If only a subset of experiments are reproducible, they should state which ones are omitted from the script and why.
        \item At submission time, to preserve anonymity, the authors should release anonymized versions (if applicable).
        \item Providing as much information as possible in supplemental material (appended to the paper) is recommended, but including URLs to data and code is permitted.
    \end{itemize}

\item {\bf Experimental setting/details}
    \item[] Question: Does the paper specify all the training and test details (e.g., data splits, hyperparameters, how they were chosen, type of optimizer, etc.) necessary to understand the results?
    \item[] Answer: \answerYes{} 
    \item[] Justification: We present dataset construction and all experimental details, such as hyperparameter settings and other experimental specifics, in Appendix \ref{appendix: dataset-details}, Appendix \ref{appendix: Reasoning Chain Data Marking}, Appendix \ref{appendix: Reasoning Chain Data Editing}, and Appendix \ref{appendix: Hallucination Detection Methods}.
    \item[] Guidelines:
    \begin{itemize}
        \item The answer NA means that the paper does not include experiments.
        \item The experimental setting should be presented in the core of the paper to a level of detail that is necessary to appreciate the results and make sense of them.
        \item The full details can be provided either with the code, in the appendix, or as supplemental material.
    \end{itemize}

\item {\bf Experiment statistical significance}
    \item[] Question: Does the paper report error bars suitably and correctly defined, or other appropriate information about the statistical significance of the experiments?
    \item[] Answer: \answerYes{} 
    \item[] Justification: The vast majority of experiments in this article report variance measurements.
    \item[] Guidelines:
    \begin{itemize}
        \item The answer NA means that the paper does not include experiments.
        \item The authors should answer ``Yes'' if the results are accompanied by error bars, confidence intervals, or statistical significance tests, at least for the experiments that support the main claims of the paper.
        \item The factors of variability that the error bars are capturing should be clearly stated (for example, train/test split, initialization, random drawing of some parameter, or overall run with given experimental conditions).
        \item The method for calculating the error bars should be explained (closed form formula, call to a library function, bootstrap, etc.)
        \item The assumptions made should be given (e.g., normally distributed errors).
        \item It should be clear whether the error bar is the standard deviation or the standard error of the mean.
        \item It is OK to report 1-sigma error bars, but one should state it. The authors should preferably report a 2-sigma error bar rather than state that they have a 96\% CI, if the hypothesis of Normality of errors is not verified.
        \item For asymmetric distributions, the authors should be careful not to show in tables or figures symmetric error bars that would yield results that are out of range (e.g., negative error rates).
        \item If error bars are reported in tables or plots, the authors should explain how they were calculated and reference the corresponding figures or tables in the text.
    \end{itemize}

\item {\bf Experiments compute resources}
    \item[] Question: For each experiment, does the paper provide sufficient information on the computer resources (type of compute workers, memory, time of execution) needed to reproduce the experiments?
    \item[] Answer: \answerYes{} 
    \item[] Justification: We report resource consumption metrics for all experimental procedures in this study.
    \item[] Guidelines:
    \begin{itemize}
        \item The answer NA means that the paper does not include experiments.
        \item The paper should indicate the type of compute workers, CPU or GPU, internal cluster, or cloud provider, including relevant memory and storage.
        \item The paper should provide the amount of compute required for each of the individual experimental runs, as well as estimate the total compute. 
        \item The paper should disclose whether the full research project required more computing than the experiments reported in the paper (e.g., preliminary or failed experiments that didn't make it into the paper). 
    \end{itemize}
    
\item {\bf Code of ethics}
    \item[] Question: Does the research conducted in the paper conform, in every respect, with the NeurIPS Code of Ethics \url{https://neurips.cc/public/EthicsGuidelines}?
    \item[] Answer: \answerYes{} 
    \item[] Justification: All aspects of this work comply with the NeurIPS Code of Ethics.

\item {\bf Broader impacts}
    \item[] Question: Does the paper discuss both potential positive societal impacts and negative societal impacts of the work performed?
    \item[] Answer: \answerYes{} 
    \item[] Justification: A detailed discussion of both positive and negative societal impacts is provided in Appendix~\ref{appendix: broader_impact}.

    \item[] Guidelines:
    \begin{itemize}
        \item The answer NA means that there is no societal impact of the work performed.
        \item If the authors answer NA or No, they should explain why their work has no societal impact or why the paper does not address societal impact.
        \item Examples of negative societal impacts include potential malicious or unintended uses (e.g., disinformation, generating fake profiles, surveillance), fairness considerations (e.g., deployment of technologies that could make decisions that unfairly impact specific groups), privacy considerations, and security considerations.
        \item The conference expects that many papers will be foundational research and not tied to particular applications, let alone deployments. However, if there is a direct path to any negative applications, the authors should point it out. For example, it is legitimate to point out that an improvement in the quality of generative models could be used to generate deepfakes for disinformation. On the other hand, it is not necessary to point out that a generic algorithm for optimizing neural networks could enable people to train models that generate Deepfakes faster.
        \item The authors should consider possible harms that could arise when the technology is being used as intended and functioning correctly, harms that could arise when the technology is being used as intended but gives incorrect results, and harms following from (intentional or unintentional) misuse of the technology.
        \item If there are negative societal impacts, the authors could also discuss possible mitigation strategies (e.g., gated release of models, providing defenses in addition to attacks, mechanisms for monitoring misuse, mechanisms to monitor how a system learns from feedback over time, improving the efficiency and accessibility of ML).
    \end{itemize}
    
\item {\bf Safeguards}
    \item[] Question: Does the paper describe safeguards that have been put in place for the responsible release of data or models with a high risk for misuse (e.g., pretrained language models, image generators, or scraped datasets)?
    \item[] Answer: \answerNA{} 
    \item[] Justification:  This work does not involve high-risk models or datasets, so no additional release safeguards are required.
    \item[] Guidelines:
    \begin{itemize}
        \item The answer NA means that the paper poses no such risks.
        \item Released models that have a high risk for misuse or dual-use should be released with necessary safeguards to allow for controlled use of the model, for example, by requiring that users adhere to usage guidelines or restrictions to access the model or implementing safety filters. 
        \item Datasets that have been scraped from the Internet could pose safety risks. The authors should describe how they avoided releasing unsafe images.
        \item We recognize that providing effective safeguards is challenging, and many papers do not require this, but we encourage authors to take this into account and make a best-faith effort.
    \end{itemize}

\item {\bf Licenses for existing assets}
    \item[] Question: Are the creators or original owners of assets (e.g., code, data, models), used in the paper, properly credited and are the license and terms of use explicitly mentioned and properly respected?
    \item[] Answer: \answerYes{} 
    \item[] Justification: Yes, the creators or original owners of all assets (e.g., code, data, models) used in this paper are properly credited. Additionally, the relevant licenses and terms of use are explicitly mentioned and fully respected.

\item {\bf New assets}
    \item[] Question: Are new assets introduced in the paper well documented, and is the documentation provided alongside the assets?
    \item[] Answer: \answerYes{} 
    \item[] Justification: Yes, all new assets introduced in the paper are thoroughly documented. The corresponding documentation is provided alongside these assets for clarity and reproducibility.

\item {\bf Crowdsourcing and research with human subjects}
    \item[] Question: For crowdsourcing experiments and research with human subjects, does the paper include the full text of instructions given to participants and screenshots, if applicable, as well as details about compensation (if any)? 
    \item[] Answer: \answerNA{} 
    \item[] Justification: This paper does not involve crowdsourcing experiments or research with human subjects, so such details are not included.

\item {\bf Institutional review board (IRB) approvals or equivalent for research with human subjects}
    \item[] Question: Does the paper describe potential risks incurred by study participants, whether such risks were disclosed to the subjects, and whether Institutional Review Board (IRB) approvals (or an equivalent approval/review based on the requirements of your country or institution) were obtained?
    \item[] Answer: \answerNA{} 
    \item[] Justification: This study did not involve human participants, so no risks, disclosures, or IRB approvals were required or obtained.

\item {\bf Declaration of LLM usage}
    \item[] Question: Does the paper describe the usage of LLMs if it is an important, original, or non-standard component of the core methods in this research? Note that if the LLM is used only for writing, editing, or formatting purposes and does not impact the core methodology, scientific rigorousness, or originality of the research, declaration is not required.

    \item[] Answer: \answerYes{} 
    \item[] Justification: The use of large language models is described in detail in both the main text and the appendix.

\end{enumerate}
\newpage
\appendix
\section{Broader impact} \label{appendix: broader_impact}
We believe this work facilitates the safer and more responsible deployment of large reasoning models by systematically addressing hallucination issues. Through extensive experimental insights and analyses, our study highlights several promising directions for mitigating hallucinations and enhancing model reliability. Our findings can guide future researchers and practitioners towards designing more robust and aligned systems by deepening the understanding of the mechanisms behind hallucinations. We do not anticipate any direct negative societal impacts arising from this research.

\textbf{Limitations}
In this work, we construct a mathematical model and conduct CoT attribution audits, trying to reveal one of the key causes of hallucinations: the RLLM's failure to assess its metacognitive confidence derived from incorrect knowledge. We uncover that incorrect knowledge can be mistakenly amplified during reflections, ultimately resulting in a hallucinated answer. While we identify instances of misplaced confidence, our findings are based on qualitative audits rather than quantitative confidence estimation. This may introduce potential bias to some extent and limit our ability to intervene during inference. In future work, we aim to investigate the underlying mechanisms of hallucination in RLLMs further, focusing on systematically modeling confidence dynamics during reflective reasoning. Additionally, we haven't explored effective hallucination mitigation strategies in black-box settings, which remains an essential direction for continued research.

\textbf{Future Work}
A natural next step is to quantitatively test the confidence update model proposed in Section~\ref{subsec: Reflection and Metacognition}. 
While our current analysis uses the model primarily as an interpretive lens, future work will focus on empirically bridging the gap between theory and measurement. 
Specifically, we plan to design experiments that approximate the variables in Equations~(2) and (3), for instance by leveraging entropy-based uncertainty, logit margins, and self-consistency scores as proxies for $\texttt{conf}(\cdot)$ at the claim level. 
These measurements would enable us to track how confidence evolves across reasoning chains, rather than treating each step in isolation.

Another direction is to systematically compare our formulation with existing hallucination detection methods, thereby clarifying how local confidence signals interact with global dynamics of belief revision. 
Such a comparison would allow us to test whether sudden shifts in confidence, as observed in Appendix~\ref{app: Additional Experiments}, can be reliably linked to downstream hallucinations. 
Ultimately, we aim to establish a more rigorous empirical pipeline for quantifying metacognitive dynamics in long-CoT reasoning, which could in turn inform the design of training strategies and evaluation metrics that directly account for confidence evolution.

\section{Details of Dataset} \label{appendix: dataset-details}

\subsection{Dataset Overview}
Our Controlled Hallucination Audit Dataset, the first to audit Long-CoT hallucinations in RLLMs, primarily comprises question and reasoning-answer generation. All data synthesis was conducted under strict human oversight to ensure annotation quality. The dataset is divided into four subsets: Type I (Seen but Unlearned), Type I Control (Correct Answer), Type II (Unseen or Erroneous), and Type II Control (Error Rejected). All RLLMs employed for question and reasoning–answer synthesis utilize DeepSeek-R1 \cite{liu2024deepseek}. 
We chose DeepSeek-R1 as the main tested model because its characteristics align well with the needs of our study. Specifically, the model’s relatively high hallucination rate provides a fertile ground for systematic error analysis, while its low inference cost enables large-scale experimentation.
In addition, its strong recognition within the open-source community ensures reproducibility and broad relevance. These properties make DeepSeek-R1 an appropriate foundation for controlled analysis of reasoning errors in CoT-style outputs (see Appendix~\ref{appendix: generalizability} for comparative results on other models).The following presents the dataset’s construction principles and workflow.

Our Controlled Hallucination Audit Dataset's core construction principles are summarized in Table~\ref{tab:principles}. Each question and embedded false fact is based on selected RFC documents and subjected to human-model joint validation to ensure fact-driven content, no misleading information, and consistency. The data synthesis process follows principles of domain confinement, template-based design, traceability, multi-round sampling with consistency checks, and metadata recording.

\begin{table}[ht]
\centering
\caption{Core construction principles for the four dataset subsets.}
\begin{tabularx}{\linewidth}{%
    >{\raggedright\arraybackslash}p{4cm}
    >{\raggedright\arraybackslash}X
  }
\toprule
Subset & Principles \\
\midrule
\rowcolor{gray!15}
Type I\newline(Seen but Unlearned) &
1. Template-based, factually correct questions from RFC. \newline
2. Questions traceable to and exclusively sourced from the RFC. \newline
3. Wrong answers to known-fact questions. \\[0.5em]
Type I Control\newline(Correct Answer) &
1. Open-ended, factually correct questions from RFC. \newline
2. Questions traceable to and exclusively sourced from the RFC. \newline
3. Correct answers to known-fact questions. \\[0.5em]
\rowcolor{gray!15}
Type II\newline(Unseen or Erroneous) &
1. Open-ended questions from RFC documents embedding false or out-of-domain facts. \newline
2. False facts modified from the RFC were introduced and documented. \newline
3. Acceptance of false facts indicates hallucinations. \\[0.5em]
Type II Control\newline(Error Rejected) &
1. Question–answer pairs from Type II that identify all false facts. \newline
2. Corrected answers to introduced-error questions.\\
\bottomrule
\end{tabularx}

\label{tab:principles}
\vspace{-1em}
\end{table}

Table~\ref{tab:workflow} outlines our dataset's overall workflow. For Type I question–answer pairs, we use a fixed prompt template to generate a set of questions for each RFC and sample multiple answers. If human–model joint checks find consistent sampling with factual errors, we classify the pair as Type I. For Type I Control pairs, we use a prompt to open-endedly generate ``why'' questions based on the RFC and sample multiple answers. If checks find no factual errors and consistent answers, we classify the pair as Type I Control. For Type II and Type II Control, we use a prompt to open-endedly generate ``why'' questions embedding three false facts and collect multiple answer samples. If checks fail to correct all three errors or any hallucination appears, we classify the pair as Type II; otherwise, we classify it as Type II Control.

\begin{table}[H]
\centering
\caption{Mainly workflow for constructing each dataset subset.}
\begin{tabularx}{\linewidth}{%
    >{\raggedright\arraybackslash}p{4cm}
    >{\raggedright\arraybackslash}X
  }
\toprule
Subset & Workflow \\
\midrule
\rowcolor{gray!15}
Type I\newline(Seen but Unlearned) &
1. For each RFC, use a unified prompt to generate template-based questions that strictly adhere to the RFC facts. \newline
2. Submit the generated questions to the RLLMs and collect multiple sampled answers.\newline
3. Under human-model supervision, compare the sampled answers against the known facts and mark questions with incorrect facts and consistent responses as hallucination instances.\\[0.5em]
Type I Control\newline(Correct Answer) &
1. For each RFC, use a prompt to generate open-ended ``Why'' questions that strictly adhere to the RFC facts. \newline
2. Sample multiple answers and verify under human–model supervision. \newline
3. Label pairs with no factual errors and consistent responses as Type I Control. \\[0.5em]
\rowcolor{gray!15}
Type II\newline(Unseen or Erroneous) &
1. Use a prompt for each RFC to generate template-based questions embedding false facts modified from the document. \newline
2. Submit the generated questions to the RLLMs and collect multiple sampled answers.\newline
3. Under human–model supervision, label pairs as Type II if answers are inconsistent across multiple samples or do not entirely correct all embedded false facts.\\[0.5em]
Type II Control\newline(Error Rejected) &
Type II Control comprises the complement of Type II, namely those question–answer pairs that fully correct all embedded false facts under human–model supervision.\\
\bottomrule
\end{tabularx}
\label{tab:workflow}
\end{table}

In the following sections, we present a detailed account of dataset construction, outlining four primary components: Type I (Seen but Unlearned), Type I Control (Correct Answer), Type II (Unseen or Erroneous), and Type II Control (Error Rejected), to document our methodology accurately.

\subsection{Assessment of Generalizability Across Models}
\label{appendix: generalizability}

Model selection was guided by three practical considerations: (i) the frequency of hallucination phenomena, especially in CoT-induced settings; (ii) accessibility and inference cost; and (iii) adoption and relevance within the open-source community. DeepSeek-R1 was ultimately chosen because it combines a relatively high hallucination rate with low inference cost and broad community recognition, making it well suited for controlled, large-scale experimentation. While our primary experiments were conducted on DeepSeek-R1, we emphasize that the observed phenomena are not unique to this model. 

Moreover, we carried out a survey of hallucination behaviors across several reasoning-capable LLMs. We report below the results of evaluations on Claude-3.7-Sonnet and Qwen3, using the same Type I / Type II setup as in our main study. These results confirm that error propagation, reflection failure, and prompt-aligned drift are not specific to DeepSeek-R1, but rather represent broader behaviors of reasoning-oriented LLMs.

\textbf{Hallucination acceptance rate.} Table~\ref{tab:accept_rate} shows the proportion of queries (3 attempts per query) where hallucination was observed at least 2 times. Results indicate that all tested models demonstrate substantial susceptibility to hallucination under both Type I and Type II conditions, although with differing control acceptance rates. The reported values for DeepSeek-R1 are calculated over the full set of generated outputs, consistent with Table~\ref{tab:dataset statistics}. By contrast, Claude-3.7-Sonnet and Qwen3 were evaluated only on subsets that had already been filtered through DeepSeek-R1’s generation pipeline, leading to different sample distributions.

\begin{table}[h]
\centering
\caption{Hallucination acceptance rates.}
\label{tab:accept_rate}
\begin{tabular}{lcccc}
\toprule
Model & Type I & Type I Control & Type II & Type II Control \\
\midrule
DeepSeek-R1 & 62.5\% & 92.6\% & 56.1\% & 11.0\% \\
Claude-3.7-Sonnet & 73.3\% & 83.3\% & 50.0\% & 93.3\% \\
Qwen3 & 100\% & 83.3\% & 63.3\% & 83.3\% \\
\bottomrule
\end{tabular}
\end{table}

\textbf{Hallucination rate across responses.} Table~\ref{tab:halluc_rate} reports the overall hallucination rate (i.e., proportion of hallucinated answers among all generated responses). Both Claude-3.7-Sonnet and Qwen3 exhibit high hallucination frequencies, reinforcing that the tendencies identified in our main experiments extend beyond a single model.

\begin{table}[h]
\centering
\caption{Hallucination rate across all generated responses.}
\label{tab:halluc_rate}
\begin{tabular}{lcc}
\toprule
Model & Type I & Type II \\
\midrule
Claude-3.7-Sonnet & 67.8\% & 52.2\% \\
Qwen3 & 94.4\% & 65.5\% \\
\bottomrule
\end{tabular}
\end{table}

Due to resource limitations, we were unable to include GPT-o3 in these comparisons, as its inference costs exceeded our budget at the time of study. Nevertheless, these results indicate that the phenomena we audit—hallucination propagation, reflection failure, and prompt-aligned bias—are not confined to DeepSeek-R1, but generalize across diverse model families. We view our work as establishing the methodology and analysis tools, which can be readily extended to additional models in future investigations.

\subsection{Type I (Seen but Unlearned) Question-Answer Generation}
\label{appendix: type1}
This phase aims to generate a set of factually correct and verifiable question–answer pairs within the controlled knowledge domain (\(d \subset W\)), to evaluate the model's ability to answer known information. We instantiate questions for the Type I (Seen but Unlearned) scenario using predefined templates, with all questions derived from selected excerpts of RFC documents to ensure domain constraints and verifiability. Both questions and reasoning–answer pairs are generated by DeepSeek-R1, and the entire process incorporates direct human supervision and human–model joint consistency checks to guarantee data quality and reliability.

\textbf{Background and Principles.}
This phase's question generation relies exclusively on RFC (Request for Comments) documents maintained by the Internet Engineering Task Force (IETF). RFCs provide formally defined network protocol specifications that are well-structured, publicly accessible, and authoritative, offering a stable technical knowledge base.

RFC facts include explicit references that allow unambiguous answer verification. Each RFC constitutes a self-contained domain that prevents external information leakage. The RFC series covers protocol definitions, mechanisms, and parameter values, supporting the creation of diverse question templates. As official Internet standards, RFCs exhibit long-term stability and industry authority, ensuring knowledge consistency over time and mitigating issues arising from gaps in the model's pretraining data coverage.

All questions are drawn from selected RFC excerpts and strictly confined to the predefined knowledge domain (\(d \subset W\)). Each question and its expected answer are based solely on factual content without errors or fabricated information. We employ fixed templates such as ''What is the publication date of RFC \{number\}?'' to guarantee structural consistency and enable large-scale automated generation. Finally, each question undergoes human review and automated consistency checks to confirm factual accuracy and answerability. 

Questions are generated in batches using these fixed templates to ensure uniform format and scalable synthesis. We perform multi-round sampling of model outputs and apply human–model joint consistency checks to filter out sporadic errors. Direct human supervision is applied throughout question–answer generation, and all audit outcomes, including source RFC number, template ID, sampling count, and review verdict, are archived to ensure traceability and reproducibility.

\textbf{Workflow.}
We prepare an index table of all 314 RFC documents, recording fields like document number, publication date, and obsoletes relationship. We parse this index to select documents with valid entries in at least one target field, such as obsoletes or publication date, as candidates for question generation. This ensures each candidate document contains the required facts and excludes those without usable entries. The pseudocode for Type I (Seen but Unlearned) Question–Answer Generation is shown in Algorithm~\ref{alg:type1}. A simplified sample example is shown in Figure~\ref{fig:type1_sample}. The question generation prompt is displayed in Figure~\ref{fig:type1_prompt}.

\begin{enumerate}[leftmargin=*, label=(\arabic*)]
  \item For each selected RFC, use a unified prompt to generate template-based questions that strictly adhere to the RFC facts.
  \item Submit the prompt to the target LLMs and generate 10 questions in one batch.
  \item For each question, sample 5 independent answers, recording the RFC number and reference location for each response.
  \item Under human–model joint supervision, compare each response against the RFC facts and count occurrences of factual errors and consistencies across samples.
  \item Classify question–answer pairs with four or more errors and consistent answers as Type I, collecting all hallucination instances into the sample pool.
  \item Archive all question–answer pairs with their RFC number, template ID, sample count, reference locations, and validation outcomes to complete the Type I subset construction.
\end{enumerate}

\begin{algorithm}[ht]
\caption{Type I Subset Construction}
\label{alg:type1}
\begin{algorithmic}[1]
\Require RFC set $D$ of size 314, error threshold $t=4$, samples per question $n=5$
\Ensure Hallucination set $H$
\State Initialize $H \leftarrow \emptyset$
\ForAll{$r \in D$}
  \State $p \leftarrow \text{build\_prompt}(r)$
  \State $Q \leftarrow \text{LLM.generate\_questions}(p, 10)$
  \ForAll{$q \in Q$}
    \State $A \leftarrow \text{LLM.sample\_answers}(q, n)$
    \State recordResponses$(r, q, A)$
    \State $c \leftarrow \text{check\_consistency}(A)$
    \State $e \leftarrow \text{count\_errors}(A)$
    \If{$c = \text{true} \,\land\, e \ge t$}
      \State $H.\text{add}(r, q, A)$
    \EndIf
  \EndFor
\EndFor
\State \textbf{archive}($H$) with metadata $(\text{RFC}, \text{template\_id}, n, \text{refs}, e, c)$
\end{algorithmic}
\end{algorithm}

\subsection{Type I Control (Correct Answer) Question-Answer Generation}
This phase generates hallucination-free question–answer pairs for the correct-answer control subset. We use an open-ended prompt for each selected RFC document to generate ``Why'' questions covering the same in-domain facts as Type I. We then sample multiple answers for each question using DeepSeek-R1, recording each response with its RFC reference. Under human–model joint supervision, all sampled answers are verified for factual accuracy and consistency, and only pairs passing both checks are retained as Type I Control.

\textbf{Background and Principles.}
This phase uses a new set of 50 RFC documents. We use an open‐ended prompt to generate ``Why'' questions that probe in‐domain facts and ensure question diversity, with each question and its expected answer strictly fact‐driven. We sample multiple answers with DeepSeek‐R1 and verify them through human–model joint supervision to ensure answer correctness and filter out hallucinations. We record the RFC number, prompt details, consistency results, and validation outcomes for each hallucination-free pair.

\textbf{Workflow.}
The pseudocode for Type I Control (Correct Answer) Question–Answer Generation is shown in Algorithm~\ref{alg:type1c}. A simplified sample example is shown in Figure~\ref{fig:type1c_sample}. The question generation prompt is shown in Figure~\ref{fig:type1c_prompt}.

\begin{enumerate}[leftmargin=*, label=(\arabic*)]
  \item For each candidate RFC, use a unified prompt to open-endedly generate 10 fact-based, error-free ``Why'' questions.
  \item For each question, sample 5 independent answers and collect all responses.
  \item Under human–model joint supervision, verify that the 5 answers are consistent and contain no factual errors.
  \item Select the question–answer pairs that pass these checks and label them as Type I Control.
  \item Archive all Type I Control pairs with their RFC number, prompt ID, sample count, and validation outcomes.
\end{enumerate}

\begin{algorithm}[ht]
\caption{Type I Control Subset Construction}
\label{alg:type1c}
\begin{algorithmic}[1]
\Require RFC set $D$ of size 50, samples per question $n=5$
\Ensure Correct-answer set $C$
\State Initialize $C \leftarrow \emptyset$
\ForAll{$r \in D$}
  \State $p \leftarrow \text{build\_open\_prompt}(r)$
  \State $Q \leftarrow \text{LLM.generate\_questions}(p, 10)$
  \ForAll{$q \in Q$}
    \State $A \leftarrow \text{LLM.sample\_answers}(q, n)$
    \State $\text{recordResponses}(r, q, A)$
    \State $c \leftarrow \text{check\_consistency}(A)$
    \State $e \leftarrow \text{count\_errors}(A)$
    \If{$c = \text{true} \,\wedge\, e = 0$}
      \State $C.\text{add}(r, q, A)$
    \EndIf
  \EndFor
\EndFor
\State \textbf{archive}($C$) with metadata $(\text{RFC}, \text{prompt\_id}, n, \text{refs}, c, e)$
\end{algorithmic}
\end{algorithm}

\subsection{Type II (Unseen or Erroneous) and Type II Control (Error Rejected) Question-Answer Generation}
This phase evaluates the model's ability to detect and reject false knowledge by embedding three incorrect facts into ``Why'' questions. When the model generates knowledge units that appeared in its training corpus but are factually incorrect, it may trigger Type II hallucinations, reflecting its inability to assign near-zero confidence to false knowledge. In this phase, we construct both the hallucination subset (Type II) and the control subset (Type II Control) by using multiple sampling and human–model joint validation to distinguish question–answer pairs that fail to fully correct the embedded errors from those that successfully reject them.

\textbf{Background and Principles.}
Leveraging the same 50 RFC documents as Type I Control, we use open-ended prompts to generate ``Why'' questions embedding three intentionally introduced factual errors adapted from correct RFC content, testing the model's ability to reject and correct false knowledge. Each question is verified against the RFC via RAG retrieval to ensure it contains exactly three such errors. Questions and answers are synthesized by DeepSeek-R1, and multiple answers are sampled. Under human–model joint supervision, we verify response consistency and factual correction to distinguish Type II (fails to correct all errors) from Type II Control (successfully rejects and corrects all errors).

\textbf{Workflow.}
We prepare the index table of 50 RFC documents. We use an open-ended prompt for each selected RFC to generate ``Why'' questions, embedding three introduced errors, and sample multiple answers via DeepSeek-R1. We then apply human–model joint validation to classify each pair into Type II or Type II Control. The pseudocode for Type II Question–Answer Generation is shown in Algorithm~\ref{alg:type2}, and a simplified example appears in Figure~\ref{fig:type2_sample}. The question generation prompt is shown in Figure~\ref{fig:type2_prompt}.

\begin{algorithm}[ht]
\caption{Type II Subset Construction}
\label{alg:type2}
\begin{algorithmic}[1]
\Require RFC set $D$ of size 50, samples per question $n=5$, failure threshold $t=4$
\Ensure Hallucination set $H_2$, Control set $C_2$
\State Initialize $H_2 \leftarrow \emptyset$, $C_2 \leftarrow \emptyset$
\ForAll{$r \in D$}
  \State $p \leftarrow \text{build\_error\_prompt}(r)$
  \State $Q \leftarrow \text{LLM.generate\_questions}(p, 10)$
  \ForAll{$q \in Q$}
    \If{$\neg\,\text{check\_question}(q)$} 
      \State \textbf{continue}
    \EndIf
    \State $A \leftarrow \text{LLM.sample\_answers}(q, n)$
    \State $\text{recordResponses}(r, q, A)$
    \State $f \leftarrow \text{count\_failures}(A)$
    \If{$f \ge t$}
      \State $H_2.\text{add}(r, q, A)$
    \Else
      \State $C_2.\text{add}(r, q, A)$
    \EndIf
  \EndFor
\EndFor
\State \textbf{archive}($H_2, C_2$) with metadata $(\text{RFC}, \text{prompt\_id}, n, \text{refs}, f)$
\end{algorithmic}
\end{algorithm}

\begin{enumerate}[leftmargin=*, label=(\arabic*)]
  \item For each RFC, use a prompt to generate 10 ``Why'' questions embedding three false facts.
  \item Validate each question with RAG retrieval against the RFC to ensure it contains all three introduced errors; discard any that fail.
  \item For each validated question, sample 5 independent answers from the model.
  \item Prompt the model to check whether each answer corrects all three errors.
  \item For the hallucination set (Type II), select question–answer pairs where at least four answers fail to correct the errors.
  \item For the control set (Type II Control), select pairs where all five answers fully correct the errors.
\end{enumerate}
\vspace{-1em}

\subsection{Comparative Keyword Distribution in Long-CoT and Answers}
\begin{figure}[H]
  \centering
  \begin{minipage}{0.48\linewidth}
    \centering
    \includegraphics[width=\linewidth]{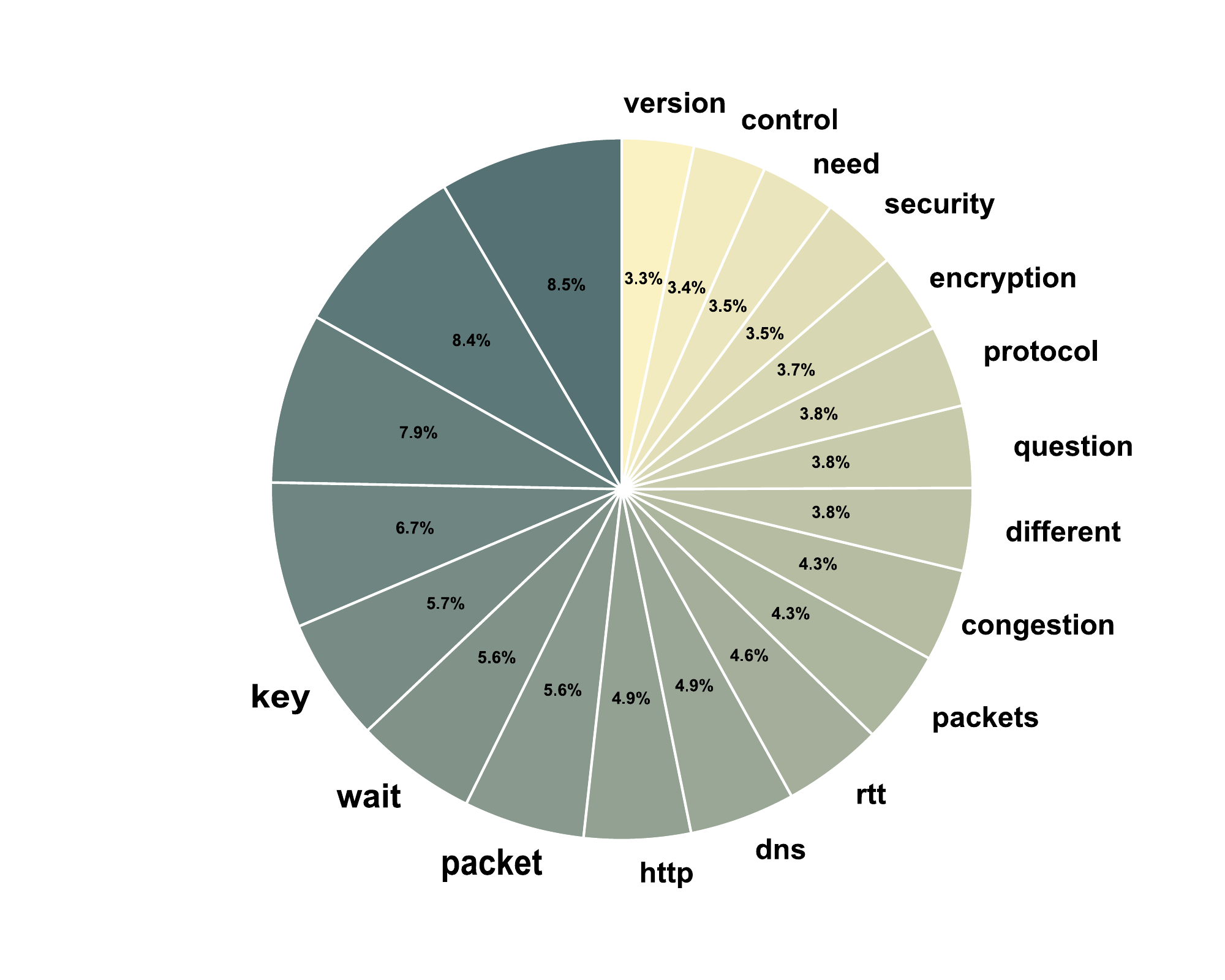}
    \captionof{figure}{Pie chart of keyword frequency in Long–CoT outputs}
    \label{fig:cot}
  \end{minipage}%
  \hfill
  \begin{minipage}{0.48\linewidth}
    \centering
    \includegraphics[width=\linewidth]{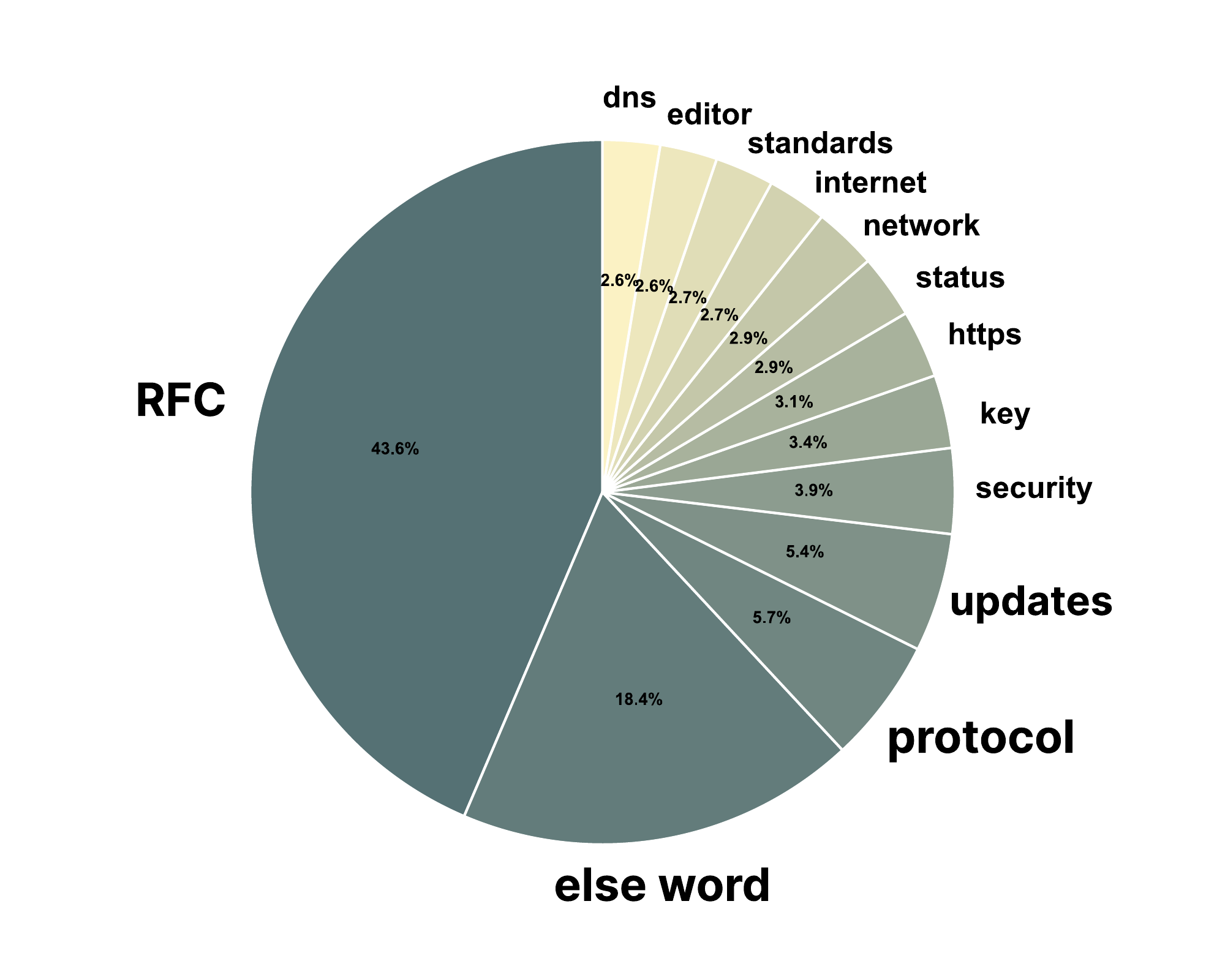}
    \captionof{figure}{Pie chart of keyword frequency in final answers}
    \label{fig:answer}
  \end{minipage}
\end{figure}

We confirm dataset adherence to RFC domain vocabulary by generating pie charts of the most frequent keywords in the Long–CoT reasoning chains (Fig.~\ref{fig:cot}) and the final answers (Fig.~\ref{fig:answer}), with all terms drawn from RFC terminology such as ``protocol'', ``header'', ``handshake'', and ``port'', indicating that both reasoning chains and answers remain tightly grounded in RFC facts and validating our template-based generation and human–model joint verification process for producing a faithful, verifiable dataset.

\subsection{Correlation Between Hallucination Frequency and CoT Length}
\label{appendix:cot_length}

To examine the relationship between hallucinations and chain length, we analyze the Type~I dataset (Appendix~\ref{appendix: type1}), where each query was answered with five independent runs. Samples were grouped by the number of hallucinations observed, and we computed the average chain-of-thought (CoT) length, measured as the number of claims, for each group.

\begin{table}[h]
\centering
\caption{Average CoT length (number of claims) under different hallucination frequencies, computed from Type~I samples with 5 independent runs.}
\label{tab:cot_length}
\begin{tabular}{lcccccc}
\toprule
Hallucinations (out of 5) & 0 & 1 & 2 & 3 & 4 & 5 \\
\midrule
Avg. CoT Length (claims) & 26.10 & 47.61 & 42.30 & 44.57 & 50.09 & 53.31 \\
\bottomrule
\end{tabular}
\end{table}

As shown in Table~\ref{tab:cot_length}, higher hallucination frequency is consistently associated with longer reasoning chains. This evidence substantiates \textbf{Obs.~II}~\ref{subsec: behavioral analysis}, namely that longer chains reflect increased reflection arising from metacognitive revision.

\subsection{Evidence for Prompt-Aligned Bias}
\label{appendix:prompt_bias}

To further substantiate our interpretation of prompt-aligned bias, we conducted an additional experiment designed to test whether the model can reliably distinguish factual from non-factual statements. We constructed a balanced evaluation set of 500 factually correct and 500 factually incorrect statements (drawn from the same source pool as the Type~II setup) and asked the model to judge their correctness under a neutral prompt.

\begin{table}[h]
\centering
\caption{Model judgments of factual correctness on 1,000 balanced statements.}
\label{tab:knowledge_test}
\begin{tabular}{lcc}
\toprule
 & Judged as Correct & Judged as Incorrect \\
\midrule
True Statements  & 478 & 22 \\
False Statements & 13  & 487 \\
\bottomrule
\end{tabular}
\end{table}

As shown in Table~\ref{tab:knowledge_test}, the model correctly classifies the majority of true and false statements, suggesting that its failures cannot be attributed to simple knowledge unavailability. Moreover, in Type~II (Unseen or Incorrect) cases selected for analysis, we did not observe any signs of the model expressing uncertainty or epistemic hesitation about the injected incorrect information (in answer). The model confidently accepted and followed the external incorrect knowledge, despite clearly “knowing better” in isolation. Taken together, this evidence supports our interpretation that the model’s behavior is not simply caused by a lack of knowledge. Instead, we argue that it reflects a prompt-aligned bias, where the model over-prioritizes consistency with the input prompt, even at the expense of factual correctness.
This reinforces our conclusion that prompt-aligned bias, rather than knowledge limitations, drives Type~II hallucinations.

\section{Details of Behavioral Analysis of Hallucinations in Long-CoT} \label{appendix: Reasoning Chain Data Marking}
This appendix corresponds to Section 3.2 and provides an overview of the complete quantitative analysis procedure performed on Type I and Type II hallucination cases using Deepseek-R1 responses. The analysis focuses exclusively on the chain of thought (CoT)—the answer is only used as contextual input. All model judgments were conducted under human supervision via the GPT-4o-mini API. The following steps constitute the full procedure for the Behavioral Analysis of Hallucinations in Long-CoT:

\begin{enumerate}
  \item \textbf{Claim Segmentation}  
    Split each CoT into individual \emph{claims}.
  
  \item \textbf{Sentence-Level Hallucination Annotation}  
    For each claim in Type I and Type II samples, mark it as a \emph{hallucinated claim} or not. Prompt: Figure~\ref{fig:1_prompt}.
  
  \item \textbf{Accepted/Corrected/Rejected Determination}  
    For every hallucinated claim, evaluate independently whether it is \emph{accepted}, \emph{corrected}, and \emph{rejected} in the full CoT. Type I uses the RFC index as context; Type II also includes three external wrong facts. Prompt: Figure~\ref{fig:2_prompt}.
  
  \item \textbf{Important Hallucinated Claims Extraction}  
    Based on the \emph{question}, CoT, \emph{answer}, and \emph{eval\_answer} (human–model agreement score), select up to five \emph{important hallucinated claims}—those whose removal or correction would significantly alter the final answer or overall reasoning—and count their repetition frequency. Type II also includes the three external wrong facts. Prompt: Figure~\ref{fig:3_prompt}.
  
  \item \textbf{Reflection Times Counting}  
    Using the \emph{question}, CoT, and \emph{answer}, count the total \emph{reflection times}—instances where the model self-evaluates its reasoning. Prompt: Figure~\ref{fig:4_prompt}.
\end{enumerate}

\subsection{Annotation Pipeline Criteria}
\label{appendix: annotation}
To ensure annotation quality and consistency, we adopted a rigorous, multi-stage pipeline that combines GPT-4o-assisted tagging with human verification. In particular, as described in Appendix~\ref{appendix: more_cases}, we defined the following categories with precise criteria:
\begin{itemize}
    \item \textbf{Wrong Reasoning}. This refers to a sentence or group of sentences responsible for “drawing conclusions or summarizing” within the reasoning chain, but which ultimately arrives at a judgment or answer that is clearly inconsistent with the facts. In simple terms, the model continues reasoning based on an incorrect premise and incorrectly accepts it.
    \item \textbf{External Incorrect Knowledge}. This refers to a sentence or group of sentences in which the model references or builds upon external knowledge introduced directly or indirectly by the user input (i.e., information not contained in the model’s internal knowledge base or the relevant RFC document). These statements contain factual errors because the model accepts, incorporates, or elaborates on user-supplied information that is itself incorrect or misleading. In short, the model incorrectly relies on “imported” knowledge provided through the prompt.
    \item \textbf{Internal Incorrect Knowledge}. This refers to fact-based content produced by the model that stems from its own internal knowledge, not prompted or introduced by the user. The model treats this information as objective truth, often presenting it with confidence, but it is factually incorrect when checked against the authoritative RFC document. In short, it reflects mislearned or misremembered knowledge from the model’s prior training or internal reasoning.
    \item \textbf{Unreasonable Assumptions}. This refers to unsupported, disconnected assumptions raised by the model in its reasoning, often introduced with conditional language such as “if…” or “suppose…”. These assumptions lack justification from the context or facts, leading to a flawed logical foundation from the outset.
    \item \textbf{Self-queries}. This refers to rhetorical or reflective questions posed by the model to itself during reasoning, often to explore or test new ideas. These typically end in question marks or include phrases like “let me think,” “could it be,” or “wait…,” guiding the model’s next steps.
\end{itemize}

Following GPT-4o–based annotation, we manually sampled 10\% of the annotated dataset to refine the labeling schema, correcting edge cases and ambiguous boundaries between categories. The final annotation pipeline used in the study is the result of multiple iterations of refinement and validation. 

\subsection{Explanation of three types of CoT trajectory}
\label{appendix: Explanation of Figure 2}
Here we present three real-world cases in Figure~\ref{fig: case 1} to provide a clearer explanation.
Figure~\ref{fig:case1(a)} illustrates a Type I case where no external errors are introduced, yet the model spontaneously generates incorrect internal knowledge (e.g., $ck_1$).
Some of these, such as $ck_1$, are assigned low confidence, downgraded into self-queries, and then correctly dropped.
In another branch, the model briefly reaches the correct claim $c_5$ via reflection but later drops it at the next wrong reasoning step.
Notably, the propagation of $ck_4$ causes the model to amplify a previously low-confidence self-query claim ($c_9$) through an incorrect reflection by mistake. 
This incorrect reinforcement leads to overconfident metacognitive judgment on false information, ultimately resulting in an incorrect final answer.
Figure~\ref{fig:case1(b)} serves as a control case, illustrating how the model successfully rejects injected incorrect external knowledge(e.g., $ck_1$). These claims are introduced early but are identified and effectively dropped during subsequent reasoning.
Even though an internal error ($ck_5$) emerges later in CoT, it is corrected through proper reflection, allowing the model to arrive at the correct final answer.
Figure~\ref{fig:case1(c)}(c) presents a complete Type II case. While with the incorrect external knowledge ($ck_3$) added, the model generates non-hallucinated claims ($c_3$) based on its own knowledge ($ck_1$, $ck_2$), the misleading nature of the prompt significantly impacts its ability to make judgments.
Despite partial reflection attempts (e.g., $c_5 \rightarrow c_6$, $c_7 \rightarrow c_8$), the model appears to self-persuade---likely due to strong instruction-following ability or prompt alignment.
It constructs hypothetical assumptions (e.g., $if$) to force an unreasonable explanation, which further induces incorrect internal knowledge ($ck_4$), ultimately leading to a hallucinated final answer.

\subsection{Case Study}\label{appendix: more_cases}
Below, we present real-world examples extracted from \textbf{\textit{Type I}} and \textbf{\textit{Type II}} cases for illustration and analysis. The highlighted colors indicate different semantic roles in the reasoning process:
\wrc{light pink for wrong reasoning},
\ek{red for external incorrect knowledge},
\ik{dark red for internal incorrect knowledge},
\wa{khaki for unreasonable assumptions},
\rf{green for reflections},
\re{yellow for task restatements}, and
\sq{blue for self-queries}.

In the \textbf{\textit{Type I}} case, after receiving the user instruction, the model recalls background knowledge related to RFC documents. However, during the reasoning process, it incorrectly introduces false RFC information. For instance, RFC 3866 is mistakenly described as ``Modifications to the Message-ID and Resent-Header Fields in the Internet Message Format'' while its actual title is ``Language Tags and Ranges in the Lightweight Directory Access Protocol (LDAP)''.
As the model continues to search for relevant RFCs, its confidence gradually declines, shifting from declarative statements to self-queries.
Ultimately, the model concludes that no RFC document is obsoleted by RFC 3866, which is factually incorrect.

\begin{tcolorbox}[sharp corners, breakable, colframe=DeepBlue, colback=white, 
       boxrule=3pt, boxsep=0.5pt, enhanced, 
       shadow={3pt}{-3pt}{0pt}{opacity=1,mygrey},
       title={Case Study (\textit{\textbf{Type I}})},]\label{box:operator-profile}
       \footnotesize
       {
\textbf{Question:} Please tell me which RFC was obsoleted by RFC 3866.\\

\textbf{CoT:} \re{Okay, so I need to figure out which RFC was obsoleted by RFC 3866.} Let me start by recalling what RFCs are. RFC stands for Request for Comments, and they are documents published by the Internet Engineering Task Force (IETF) that describe methods, behaviors, research, or innovations applicable to the working of the Internet and Internet-connected systems. Now, when an RFC is obsoleted, it means that a newer RFC has replaced it, either by updating the content, correcting errors, or expanding on the original. \re{So RFC 3866 must have made an older RFC obsolete. To find out which one, I can check the header of RFC 3866.}

...

\ik{I know that RFC 3866 is related to the modification of the Message-ID and Resent-* fields in email headers. Specifically, it updates the standards for generating Message-IDs in emails to ensure they are globally unique.} \rf{But I think there was another RFC that specifically dealt with the Message-ID requirements.} \ik{Wait, RFC 1036 is related to USENET messages, } but maybe that's not it. \ik{Alternatively, RFC 822 was the original for ARPA Internet text messages, which included email headers.} \sq{Wait, maybe RFC 3866 obsoleted RFC 822 in part regarding the Message-ID? Or perhaps a different RFC?}

...

\ik{Wait, actually, looking back, RFC 3866 is titled ``Modifications to the Message-ID and Resent-* Header Fields in the Internet Message Format''.} \wrc{So it modifies the rules for these fields, which were previously defined in RFC 2822.} \sq{Therefore, if RFC 3866 updates part of RFC 2822, does it obsolete RFC 2822?} \ik{No, because RFC 2822 was later obsoleted by RFC 5322.} \wrc{But perhaps RFC 3866 obsoletes an older RFC that specifically dealt with Message-IDs. Wait, perhaps RFC 822. Let me think: RFC 822 was obsoleted by RFC 2822, which was then obsoleted by RFC 5322. So RFC 3866 might have obsoleted an earlier RFC that dealt with Message-ID syntax.} \sq{Wait, perhaps RFC 1036?} Let me check. RFC 1036 is titled ``Standard for interchange of USENET messages'', which does define Message-ID but for USENET. However, email and USENET have different standards. \sq{Does that right?}

...

Wait, let me check. \re{The user's question is about RFC 3866.} \ik{If I search ``RFC 3866 obsoletes'', the top result is from the RFC Editor page which states: ``Network Working Group C. Karpilovsky Request for Comments: 3866 SBC Internet Services Updates: 2822 July 2004 Category: Standards Track Modifications to the Message-ID and Resent-* Header Fields in the Internet Message Format''} \wrc{So indeed, it only updates 2822.}

...

\wrc{Therefore, the answer is that there is no RFC obsoleted by RFC 3866. However, if the user insists on an answer, perhaps there's a mistake in the question. But given the information available, the accurate response is that RFC 3866 does not obsolete any RFC but updates RFC 2822.}

       }
\end{tcolorbox}

In the \textbf{\textit{Type II}} case, the model initially lists correct internal knowledge relevant to the question, such as the definitions of SRH and the UDP checksum. 
However, once incorrect knowledge is introduced into the reasoning process, the model is  misled successfully. 
This triggers a cascade of additional internal errors, with more and more incorrect internal knowledge being incorporated into the reasoning process.
Throughout this process, it repeatedly attempts to return to the key point that the UDP checksum ensures integrity rather than security through reflections.
Yet, due to prompt alignment and self-persuasion, the model ultimately fails to correct its wrong reasoning, resulting in a hallucinated answer.

\begin{tcolorbox}[sharp corners, breakable, colframe=DeepBlue, colback=white, 
       boxrule=3pt, boxsep=0.5pt, enhanced, 
       shadow={3pt}{-3pt}{0pt}{opacity=1,mygrey},
       title={Case Study (\textit{\textbf{Type II}})},]\label{box:operator-profile}
       \footnotesize
       {
\textbf{Question:} Why must SRH nodes validate UDP checksums to prevent HMAC spoofing per Section 5.2?\\

\textbf{``wrong\_fact1'':} ``UDP checksum validation is tied to HMAC security.''\\  
\textbf{``wrong\_fact2'':} ``Section 5.2 links transport-layer checksums to SRH integrity.''\\  
\textbf{``wrong\_fact3'':} ``HMAC validation requires UDP checksum verification.''\\  

\textbf{CoT:}
\re{Okay, so I need to figure out why SRH nodes must validate UDP checksums to prevent HMAC spoofing according to RFC Section 5.2. Let me start by breaking down the terms here.}  
First, SRH probably stands for Segment Routing Header, which is a type of routing extension header in IPv6. Segment Routing is a network architecture that enhances packet routing by specifying a path through segments. The SRH would contain information about the path a packet should take.  
Next, UDP checksums. UDP is the User Datagram Protocol, a transport layer protocol. The checksum in UDP is used to detect errors in the packet. When a UDP packet is sent, the checksum is computed and included. The receiver recalculates it to check for errors. If the checksum doesn't match, the packet is usually discarded.  
\ek{Now, the question is why validating the UDP checksum is necessary to prevent HMAC spoofing. The RFC section 5.2 probably discusses security considerations for SRH, specifically regarding HMAC validation.}  
\wrc{Putting this together: If a node doesn't validate the UDP checksum, maybe an attacker could tamper with the UDP payload (which might include the HMAC) without the checksum catching it. Since the checksum is a way to ensure data integrity, not validating it would allow the payload to be altered undetected.}  
\wrc{If the HMAC is part of the UDP payload, then without checksum validation, an attacker could modify the HMAC (spoof it) without the node realizing, because the checksum wasn't checked.}  
\ik{The node would then process a packet with a forged HMAC, leading to potential security issues like unauthorized path changes.}  
\rf{But wait, HMAC itself is supposed to provide authentication. So why would the UDP checksum matter?}  
\ik{Perhaps the HMAC in SRH is transported within the UDP payload,}  
\wa{and if the UDP checksum isn't validated,}  
\wrc{ the node the integrity of the payload, making it easier for an attacker to alter the HMAC without detection}  
\rf{However, HMAC should already include a secret key, so altering it without knowing the key should make it invalid.}  
\ik{But perhaps in the context of SRH, the HMAC validation process relies on the UDP checksum being correct to ensure that the data used to compute the HMAC hasn't been tampered with during transit.}  
\wa{If the UDP checksum isn't checked, }  
\wrc{the node might process a corrupted packet, including a corrupted HMAC, which could lead to spoofing if other validations are bypassed. }  
\ik{Alternatively, maybe the HMAC is part of the SRH, not the UDP payload, but the UDP checksum covers the entire UDP payload, which includes the SRH.}  
\wa{If the checksum isn't validated,}  
\wrc{an attacker could modify the SRH (including the HMAC) without the node detecting the change via checksum. The node might then accept the modified SRH because the checksum wasn't checked, even though the HMAC is invalid.}  
\rf{But that seems contradictory because the HMAC should be validated separately. }  

...

Putting it all together, the RFC likely mandates UDP checksum validation to ensure that the UDP payload,  
\ik{which includes the SRH and its HMAC, hasn't been tampered with during transmission. If the checksum isn't validated, an attacker could alter the HMAC or other parts of the SRH, and even though the HMAC is a cryptographic check, there might be scenarios where not validating the checksum could allow certain spoofing attacks, especially if the HMAC validation has any dependencies on the UDP header fields that could be altered without checksum validation. By ensuring the checksum is valid, the node can trust that the UDP payload (including HMAC) hasn't been corrupted, thereby preventing HMAC spoofing attempts that rely on undetected payload alterations.}  
}
\end{tcolorbox}

\section{Details of Impact of Upstream Reasoning on Downstream Fidelity}
\label{appendix: Reasoning Chain Data Editing}
Here we describe in detail how we implemented the impact of upstream reasoning on downstream fidelity, starting with the method used to locate the First Incorrect Knowledge node in the original Chain of Thought. We then explain the three intervention points, Before First Hallucination, At First Hallucination, and After First Hallucination, and describe how we inject the corresponding correction assertion at each point. Finally, we outline the manual annotation protocol for metrics M1 to M6, covering acceptance rate, CoT alteration rate, answer alteration rate, CoT–answer consistency, propagation rate, and hallucination persistence rate. We detail our data aggregation procedures to ensure full reproducibility.

\subsection{Locating the First Incorrect Knowledge}
We first locate the first incorrect knowledge node in each chain of thought to investigate how upstream reasoning errors affect downstream answer fidelity. This step underlies the intervention experiments and allows us to assess how injecting correction assertions at different points alters the final answer.

In this experiment, we select 70 samples for validation, including 40 hallucination samples and 30 non-hallucination samples. This ensures coverage of typical error behaviors and provides a control group for comparison.

All input fields are listed below. Fields in parentheses are optional and are included only when present in the corresponding sample type:
\begin{itemize}
  \item \texttt{question}: the original question;
  \item \texttt{question\_evaluation}: evaluation of whether external wrong facts were introduced in the question;
  \item \texttt{rag\_reference}: retrieved reference passages for the question;
  \item \texttt{wrong\_facts}: content of the external wrong facts;
  \item \texttt{cot}: the full Chain-of-Thought generated by the model;
  \item \texttt{answer}: the model's final answer;
  \item \texttt{eval\_answer}: preliminary evaluation of answer correctness.
\end{itemize}

We use the ChatGPT-o3 API to run the prompt shown in Figure~\ref{fig:first_prompt} and keep only those samples whose output is identical across five runs to ensure stability and accuracy. The prompt returns the complete sentence where Incorrect Knowledge first appears in the CoT. This sentence serves as the reference point for subsequent injections. All locating results are manually verified to prevent omissions or errors, providing a reliable basis for the three intervention strategies.

\subsection{Inserting Corrective Knowledge at Intervention Points}
To systematically evaluate the effect of injecting correction assertions at different times on downstream reasoning and answer fidelity, we perform independent interactions with each sample using the ChatGPT-o3 API, simulating the model's thinking tone (e.g., ``Hmm…'') and flexibly adjusting phrasing according to the injection point to integrate the corrective information naturally into the context. The procedure is as follows:

\begin{itemize}
  \item \textbf{Before First Hallucination}  
    Insert the correction assertion at the appropriate position before the First Hallucination to steer the model away from the erroneous branch early.  
  \item \textbf{At First Hallucination}  
    Insert the correction assertion immediately at the position of the First Hallucination to provide an instant correction.  
  \item \textbf{After First Hallucination}  
    Insert the correction assertion at the appropriate position after the First Hallucination to simulate the model's reflective reconsideration.  
\end{itemize}

After inserting the assertion, we truncate the original CoT at the injection point and invoke Deepseek-R1-14B to continue generating the remaining reasoning, making the new assertion the starting point for downstream inference.

For the 30 non-hallucination samples, we manually select an early insertion point in each CoT, inject a manually written Incorrect Knowledge assertion, and then continue generation from that point.

In the following subsection, we describe the manual annotation protocol for six key metrics, Acceptance Rate, CoT Alteration Rate, Answer Alteration Rate, Propagation Rate, CoT–Answer Consistency, and Hallucination Persistence Rate, to document how each metric is applied in practice.

\subsection{Assessing Downstream Fidelity Across Six Indicators}
We select six indicators, adoption of the correction, CoT structure change, answer change, CoT–answer alignment, propagation of the correction, and persistence of hallucination, to cover the entire path from intervention to final output. Together, they quantify how each corrective insertion affects both intermediate reasoning and ultimate answer fidelity.

All assessments are carried out by human reviewers to capture subtle judgments that cannot be automated. For each edited Chain-of-Thought (CoT) and corresponding answer:

\begin{itemize}
  \item \textbf{Adoption Rate}: mark ``adopted'' if the correction assertion appears verbatim or is clearly integrated into the revised CoT.  
  \item \textbf{CoT Change Rate}: mark ``changed'' if any branch of the reasoning chain diverges from the original beyond the injection point.  
  \item \textbf{Answer Change Rate}: mark ``changed'' if the final answer's content or conclusion differs from the original.  
  \item \textbf{CoT–Answer Alignment}: mark ``aligned'' if the revised CoT logically supports the new answer without contradiction.  
  \item \textbf{Propagation Rate}: count downstream assertions or tokens that build on the correction and divide by total chain length.  
  \item \textbf{Hallucination Persistence}: mark ``persistent'' if the revised answer still contains factual errors.  
\end{itemize}

Each sample was independently reviewed by two experts. Any disagreements were discussed until consensus was reached, and a third expert resolved remaining conflicts. In total, we collected 150 annotated samples.

\subsection{Explanation of CoT Editing in Hallucinated Cases}
\label{appendix: Explanation of Figure 4}

Here we describe in detail how the CoT changes after intervention editing.
In Figure~\ref{fig: case3(a)}, the original unedited case is shown, where an incorrect answer is generated due to hallucination introduced by internal knowledge claims (e.g., $ck_3$). 
Despite a failed reflection on $ck_6$, the model ultimately produces the wrong final answer. 
The intervention editing results shown in Figures ~\ref{fig: case3(b)} --~\ref{fig: case3(d)} illustrate three cases, where the original $c_4$ from Figure~\ref{fig: case3(a)} is replaced with a newly inserted claim $ck_4'$, leading to different downstream reasoning trajectories. 
In Figure~\ref{fig: case3(b)}, the edit $ck_4'$ is accepted, leading to $c_5'$. Although $c_5'$ instructs the incorrect claim $c_7'$ to be dropped, some other errors remain ($ck_7'$), resulting in a partially improved but still incorrect answer. 
In contrast, Figure~\ref{fig: case3(c)} illustrates a case where the model directly rejects ($drop$) $ck_4'$ at $c_5'$. 
However, the downstream reasoning process is influenced by editing, accidentally introducing new hallucinations ($ck_6'$), and ultimately resulting in a hallucinated answer. 
In Figure~\ref{fig: case3(d)}, the model is guided by $c_5'$. 
It not only initiates further correct reasoning steps ($c_6' \rightarrow c_7'$), but also successfully corrects the internal incorrect claim $ck_5'$ through proper self-reflection ($ck_8'$), ultimately arriving at the correct answer.

\subsection{Explanation of CoT Editing in Correctly Answered Cases}
\label{appendix: Explanation of Figure 5}
We also conduct parallel experiments on correctly answered cases to add hallucination by editing CoT. 
Figure~\ref{fig: case4(a)} presents the unedited CoT trajectory, where despite the presence of hallucinated claims($c_3, c_6$), the model successfully corrects the error through proper reflection($ c_5, c_7$) and ultimately arrives at the correct answer. 
Figure~\ref{fig: case4(b)} - Figure~\ref{fig: case4(d)} illustrate three representative cases after editing the original CoT by truncating at claim $c_1$ and injecting incorrect knowledge $ck_3'$. 
Figure~\ref{fig: case4(b)} exemplifies a case where the injected incorrect knowledge $ck_3'$ is entirely accepted, altering the downstream reasoning and introducing additional factually incorrect claims ($ck_4'$). Although $c_5'$ was corrected through self-reflection($c_7'$), the incorrect claim $c_2'$ enabled by $ck_3'$ continues to propagate to the final answer. 
Figure~\ref{fig: case4(c)} shows a case where the model accepts the incorrect knowledge $ck_3'$ but drops it after a few steps of reasoning. 
Only slight changes happen in the subsequent CoT and final answer, with the overall reasoning trajectory remaining nearly identical to the original. 
Figure~\ref{fig: case4(d)} illustrates a case where the model immediately rejects the injected incorrect knowledge $ck_3'$ at $c_2'$. Although the subsequent CoT undergoes large changes, it maintains the correct final answer.

\section{Details of Hallucination Detection Methods} \label{appendix: Hallucination Detection Methods}

We benchmark hallucination risk in generated reasoning chains using seven paradigms grounded in uncertainty quantification or internal representation analysis.

\subsection{Evaluation Setting}

\textbf{Model Setting.}
For the Dataset Construction section, we used the DeepSeek-R1 API~\cite{liu2024deepseek} and ChatGPT-4o API~\cite{achiam2023gpt} to synthesize data and assist in the manual verification of samples. We evaluated performance based on DeepSeek's officially released distilled model, DeepSeek-R1-Distill-Qwen-14B~\cite{liu2024deepseek} for the Hallucination Detection section. Some detection methods required long-text understanding and processing, such as splitting full CoTs into individual claims, for which we employed DeepSeek-V3 API~\cite{liu2024deepseek} and  ChatGPT-4o API.

\textbf{Environment Setting}
All experiments were conducted on a Linux server running Ubuntu 20.04.1 LTS (kernel version 5.15.0-124-generic, x86\_64 architecture). The server is equipped with two Intel Xeon Gold 6248R 3.00 GHz processors (dual socket, 24 cores and 2 threads per socket, totaling 96 logical CPUs), 502 GiB of RAM, and two NVIDIA A100-SXM4-80GB GPUs. The system uses driver version 535.161.07 with CUDA 12.2. The software environment includes Python 3.9, PyTorch 2.2.0, and Hugging Face Transformers 4.39.3.

\subsection{Details of the Detection Method}
Using knowledge‐base methods to detect hallucinations in Long‐CoT samples is challenging, and as LLMs' reasoning capabilities improve, the pool of human trainers qualified for such labeling shrinks, making it harder to scale hallucination detection. Therefore, we selected seven representative hallucination detection methods that do not require external knowledge bases.

Each method offers a unique perspective on hallucination detection in Long‐CoT, capturing semantic uncertainty or deeper representation discrepancies. Together, they form a comprehensive framework for evaluating hallucinations in structured reasoning. Below is a summary of each approach.

\subsubsection{Logit-Based Detection}
Logit entropy measures the dispersion of probability mass across the most likely next tokens, making it effective at revealing the model's uncertainty and exposing potential factual inconsistencies. The model's output logits are first converted into token‐level perplexity, with higher perplexity indicating greater uncertainty or likelihood of hallucination. For detection, we compute top-k logit entropy by normalizing the entropy over the K most probable tokens, following Sriramanan et al.~\cite{sriramanan2024llm}, we select an optimal threshold on the validation set, and then apply that threshold during testing to flag hallucinated outputs.

\subsubsection{Attention-Based Detection}
The attention-kernel score is computed by taking the logarithm of each token's self-attention weight (i.e., the diagonal entries of the attention matrix) and averaging across tokens. This score effectively measures the model's degree of ``self-focus'' during reasoning: consistently high self-attention suggests the model is coherently building on its previous inference, whereas a weakened diagonal focus may indicate hallucination. Following the approach of~\cite{sriramanan2024llm}, we calculate the attention-kernel score layer by layer and determine, on a validation set, the optimal threshold for each layer. We apply the threshold corresponding to the layer that achieved the best validation performance for testing.

\subsubsection{Hidden-State Based Detection}
For each sample (question–answer pair), we obtain the hidden representations of each layer in teacher-forcing mode. For each layer's activation matrix, we compute the centered covariance and perform singular value decomposition (SVD), then average the logarithm of the singular values to derive that layer's hidden-state score. Higher scores indicate more complete, structured, and reliable internal representations; lower scores often correspond to degraded or incoherent features, potentially signaling hallucination. Following Sriramanan et al.~\cite{sriramanan2024llm}, we determine the optimal threshold for each layer on a balanced validation set, and during testing, we apply the threshold from the best-performing layer to detect hallucinations in new samples.

\subsubsection{HalluciNot (HDM2 model) }
The HDM2 model~\cite{paudel2025hallucinothallucinationdetectioncontext} detects hallucinations by verifying both context-specific facts and common‐knowledge statements against a fine‐grained taxonomy of LLM outputs. Given a user prompt, an optional context, and an LLM response, it first categorizes each sentence into one of four classes (context‐based, common‐knowledge, enterprise‐specific, or innocuous) using a lightweight classifier. For context‐based claims, it employs a retrieval‐augmented consistency check.  For common‐knowledge claims, it probes a frozen backbone LLM's internal representations to identify contradictions with widely accepted facts. It then produces a document‐level hallucination score and a group of token‐level scores. We took document‐level hallucination scores to evaluate hallucinations.

\subsubsection{Claim-Conditioned Perplexity (CCP)}
The CCP method quantifies the model's confidence in factual statements and can be used to detect latent hallucinations in generated text. First, the model's output is segmented into atomic factual claims, each representing an independent fact unit. We retrieve each token in a claim's probability distribution over the top-K candidates given the context. Using a natural language inference (NLI) model, we filter these candidates to retain only those that entail or contradict the original token, and compute a normalized ratio of their probabilities as the token's confidence score. We then aggregate the token-level confidence scores within each claim to obtain an overall uncertainty score for that claim. We average the uncertainty scores across all claims to evaluate an entire text. Finally, we determine an optimal threshold on a validation set to decide whether the text contains hallucinations. We follow the approach described in~\cite{fadeeva2024fact} to perform the computations.

\subsubsection{SelfCheckGPT}
SelfCheckGPT evaluates a model's self-consistency under diverse random sampling to detect latent hallucinations. Given a user prompt, it first generates a deterministic ``main'' response at temperature 0, then produces \(N\) stochastic samples at temperature 1. The main response is segmented into sentences, and each sentence is compared against all sampled outputs via multiple consistency checks, such as maximal BERTScore similarity inversion, NLI-based contradiction probability, n-gram model token scoring, and prompt-based ``Does this hold?'' queries. Each check yields a per-sentence uncertainty score, which are aggregated to form both sentence-level and response-level hallucination scores. A threshold tuned on a validation set is then applied to classify sentences or entire responses as hallucinated.We follow the procedure in~\cite{manakul2023selfcheckgpt} to perform these calculations.

\subsubsection{Semantic Entropy}
The Semantic-Entropy score assesses semantic uncertainty by measuring the entropy of the answer distribution obtained through multiple samplings of adversarial prompts. First, the generated text is segmented into atomic claims, and several interrogative prompts are automatically constructed for each claim. Each prompt is then sampled multiple times to gather all answers along with their generation probabilities. Using a bidirectional entailment model, answers are clustered by semantic consistency, and the probabilities within each cluster are summed to form a cluster-level distribution. The information entropy of this distribution is computed as the semantic uncertainty metric for that prompt. We then average the entropies of all prompts associated with the same claim to derive the claim's overall semantic entropy. Finally, an optimal entropy threshold is determined on the validation set and applied during testing to detect hallucinations. We follow the procedure in~\cite{farquhar2024detecting} to perform these calculations.

\subsection{Cost–Benefit Analysis of Detection Baselines}
\label{appendix:cost_benefit}

All LLM inferences are normalized to a single average time unit, denoted as $T_{\text{LLM}}$, which corresponds to one forward pass through a large reasoning model (e.g., DeepSeek-R1 or GPT-4o). The remaining variables are defined as follows:
\begin{itemize}[leftmargin=*]
    \item $S$: Total number of sentences to be evaluated.  
    \item $C_{\text{avg}}$: Average number of claims extracted per sentence (empirically $\approx 1.8$).  
    \item $Q$: Number of question variants generated per claim (typically 3).  
    \item $M$: Number of times each question is re-answered (typically 3).  
    \item $N$: Number of self-check samples per original CoT (typically 20).  
    \item $T_{\text{cla}}$: Inference time for a BERT-like classifier, where $T_{\text{cla}} \ll T$.  
    \item $n$: Lightweight post-inference operations (e.g., attention/statistics), where $n \ll T$.  
\end{itemize}

\textbf{Semantic Entropy.}  
This method decomposes sentences into atomic claims, generates $Q$ question variants per claim, and samples $M$ answers per variant. Its time complexity is:
\[
\mathrm{Time}\approx S \,(T + C_{\mathrm{avg}} \, Q \, M \, T) 
= S \,(T + 1.8 \times 3 \times 3 \, T) 
= S \,(T + 16.2 \, T) 
= 17.2 \, S \, T.
\]

\textbf{CCP (Claim Consistency via Prediction).}  
CCP uses the same decomposed claims but evaluates token-level prediction confidence via an NLI model. The cost is:
\[
\mathrm{Time}\approx S \cdot C_{\text{avg}} \cdot T 
= 1.8 \, S \, T.
\]

\textbf{Self-Check.}  
This method generates $N$ responses (e.g., 20) for each CoT and compares them with the original trace using NLI-based sentence matching:
\[
\mathrm{Time}\approx 20 \, T.
\]

\textbf{Medium-Cost Methods.}  
Logit Entropy, Attention Strength, and Spectral Entropy each require one full inference followed by lightweight internal analysis:
\[
\mathrm{Time}\approx T.
\]

\textbf{HDM2.}  
HDM2 applies a fine-tuned BERT classifier directly on the full CoT outputs:
\[
\mathrm{Time}\approx T_{\text{cla}}, \quad \text{where } T_{\text{cla}} \ll T.
\]
Inference is highly efficient—typically sub-second per sample on GPU.

\subsection{Additional Experiments} \label{app: Additional Experiments}
\textbf{Initial Reasoning Confusion Drives Hallucination.}
We conducted a segment‐wise analysis of CCP values over the Chain‐of‐Thought and answer phases in four conditions (Type I, Type I Control, Type II, Type II Control). Each output sequence was split into the first one‐third versus the last two‐thirds, the first one‐half versus the last one‐half, and the first two‐thirds versus the last one‐third, and we computed the mean CCP in each segment. We chose CCP because its per-claim perplexity estimates eliminate noise from different tokenizations of the same semantic content, giving a more precise measure of model confusion at each step.

The results show that CCP is consistently higher in the initial segments than in the later segments across all four conditions, indicating lower confidence and a greater tendency to explore unsupported or incorrect inferences at the start of generation. This confirms that the model is most confused during the early reasoning and answer steps when it has less contextual grounding.

Although CCP decreases in later segments, reflecting increased fluency, this may lead the model to perpetuate early mistakes, since low perplexity does not guarantee factual accuracy. Once an incorrect claim is introduced, the model can confidently build upon that flawed premise. These observations suggest that hallucination mitigation strategies focused on the initial reasoning steps may be most effective in reducing hallucination in Chain‐of‐Thought models.  

\begin{table}[h]
  \centering
  \caption{Comparison of CCP values in different segments of the generated sequence}
  \label{tab:ccp-segment-comparison}
  \begin{tabular}{l
                  rrr  
                  rrr  
                  rrr} 
    \toprule
    & \multicolumn{3}{c}{First 1/3 vs Last 2/3}
    & \multicolumn{3}{c}{First 1/2 vs Last 1/2}
    & \multicolumn{3}{c}{First 2/3 vs Last 1/3} \\
    \cmidrule(lr){2-4} \cmidrule(lr){5-7} \cmidrule(lr){8-10}
    Type
      & First & Last  & $\Delta\%$
      & First & Last  & $\Delta\%$
      & First & Last  & $\Delta\%$ \\
    \midrule
    Type I
      & 0.308 & 0.268 & -13.02\%
      & 0.352 & 0.295 & -16.10\%
      & 0.383 & 0.268 & -30.16\% \\
    Type I C
      & 0.296 & 0.278 &  -6.17\%
      & 0.329 & 0.315 &  -4.12\%
      & 0.344 & 0.278 & -19.14\% \\
    Type II
      & 0.320 & 0.267 & -16.45\%
      & 0.310 & 0.276 & -10.69\%
      & 0.304 & 0.267 & -12.21\% \\
    Type II C
      & 0.266 & 0.258 &  -2.76\%
      & 0.290 & 0.254 & -12.28\%
      & 0.273 & 0.258 &  -5.27\% \\
    \midrule
    \textbf{Average}
      & 0.297 & 0.268 & -10.01\%
      & 0.320 & 0.285 & -10.84\%
      & 0.326 & 0.268 & -17.20\% \\
    \bottomrule
  \end{tabular}
\end{table}

\begin{figure}[H]
  \centering
  \includegraphics[width=\textwidth,height=0.2\textwidth]{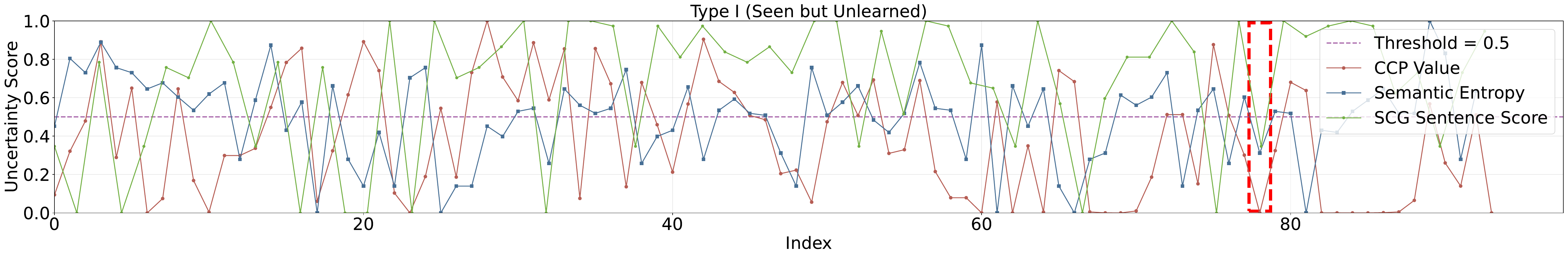}
  \captionsetup{type=figure}
  \captionof{figure}{Type I (Seen but Unlearned) Sample. The true hallucination in the red border is treated as factual.}
  \label{fig:additionalExperimentsCase1}
\end{figure}

\begin{figure}[H]
  \centering
  \includegraphics[width=\textwidth,height=0.2\textwidth]{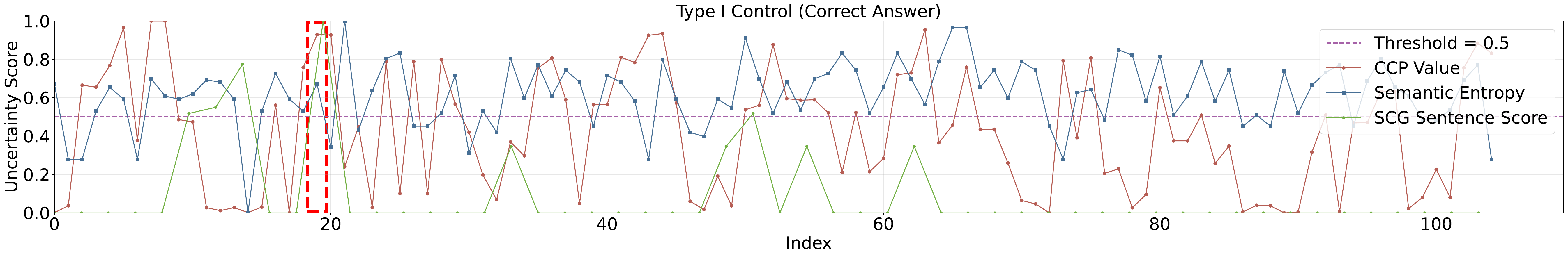}
  \captionsetup{type=figure}
  \captionof{figure}{Type I Control (Correct Answer) Sample. The hallucination-free claim in the red border is marked as hallucination.}
  \label{fig:additionalExperimentsCase2}
\end{figure}

\begin{figure}[H]
  \centering
  \includegraphics[width=\textwidth,height=0.2\textwidth]{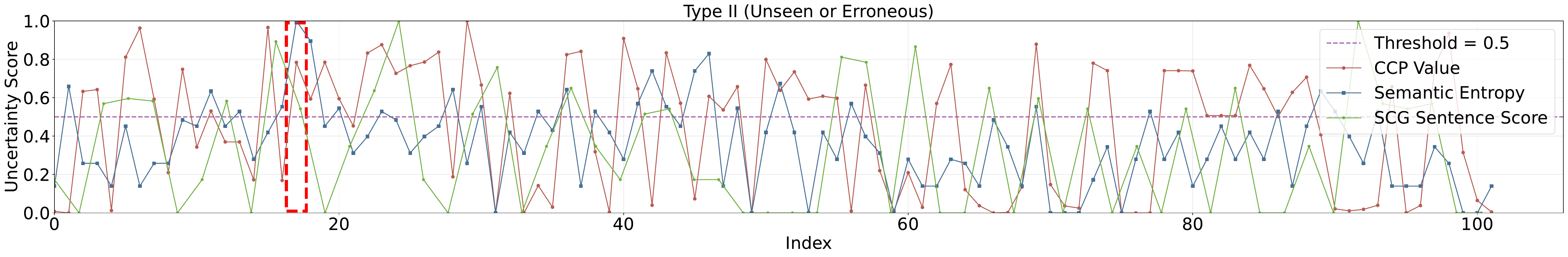}
  \captionsetup{type=figure}
  \captionof{figure}{Type II (Unseen or Erroneous) Sample. The hallucination-free claim in the red border is marked as hallucination.}
  \label{fig:additionalExperimentsCase3}
\end{figure}

\begin{figure}[H]
  \centering
  \includegraphics[width=\textwidth,height=0.2\textwidth]{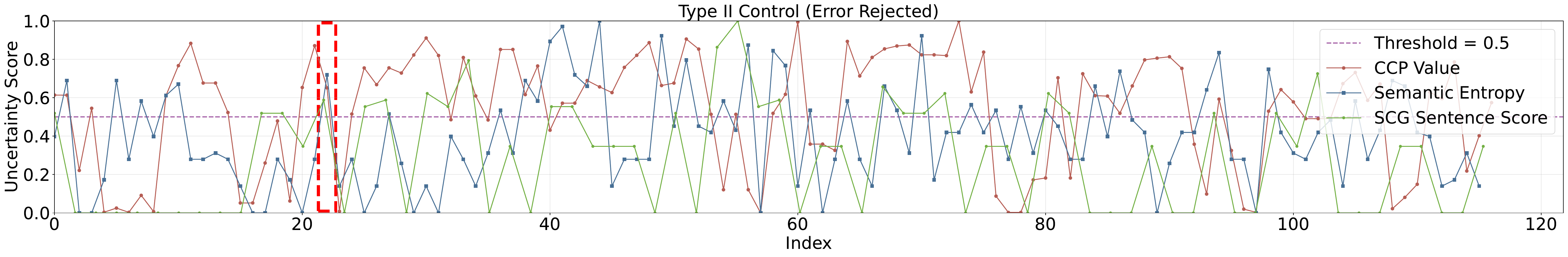}
  \captionsetup{type=figure}
  \captionof{figure}{Type II Control (Error Rejected) Sample. The true hallucination in the red border is correctly classified.}
  \label{fig:additionalExperimentsCase4}
\end{figure}

\textbf{Semantic Consistency Detectors Struggle at Claim Level.}
We perform a claim-level verification of three detectors to evaluate semantic consistency checking methods at a fine‐grained level. First, we select one representative case for each condition and plot the uncertainty score of each atomic claim (Figs.~\ref{fig:additionalExperimentsCase1}, \ref{fig:additionalExperimentsCase2}, \ref{fig:additionalExperimentsCase3}, \ref{fig:additionalExperimentsCase4}). Next, we identify the claims on which all three detectors (SelfCheckGPT, Semantic Entropy, and CCP) assign similar scores—high or all low—and manually verify whether each claim is a hallucination. Figure~\ref{fig:additionalExperimentsCase4} shows the claims correctly classified by all detectors, Figure~\ref{fig:additionalExperimentsCase1} lists those wrongly flagged as hallucinations, and Figures~\ref{fig:additionalExperimentsCase2}, \ref{fig:additionalExperimentsCase3} list genuine hallucinations that all three miss. Although each detector succeeds on some claims, a large portion are jointly misclassified: hallucination‐free claims are marked as hallucinations, and true hallucinations are treated as factual. This result indicates that methods relying on semantic consistency still have limited power to detect hallucination at the atomic‐claim granularity, and even if they exhibit some hallucination detection capability on long texts, that performance may simply reflect cancellation of sentence‐level errors rather than genuine effectiveness.

\FloatBarrier

\section{Prompt Template and Supplementary Materials} \label{appendix: prompt}

\begin{figure}[ht]
  \centering
  \begin{tcolorbox}[%
       sharp corners, breakable=false, colframe=DeepBlue, colback=white, 
       boxrule=3pt, boxsep=0.5pt, enhanced, 
       shadow={3pt}{-3pt}{0pt}{opacity=1,mygrey},
       title={Type I (Seen but Unlearned) Sample},]\label{box:operator-profile}
{\scriptsize
\begin{verbatim}
``id'': 1,
``RFC_section'': ``9030'',
``question'': ``Please introduce me the RFC 9030 in detail.'',
``question_type'': ``factually_correct'',
``answers'': [
       {
        ``answer_id'': 0,
        ``question'': ``Please introduce me the RFC 9030 in detail.'',
        ``answer'': ``RFC 9030 ...'',
        ``result'': false,
        ``eval_answer'': ``false  \n**explanation**: the provided ...'',
        ``cot'': ``Okay, so I need to ...''
       },
        (the other four answers)   
    ],
``consistent'': false,
``consistent_evaluation'': ``false\n\nthe five answers ...''
\end{verbatim}
}
  \end{tcolorbox}
  \refstepcounter{figure}
  \label{fig:type1_sample}
\end{figure}

\begin{figure}[ht]
  \centering
  \begin{tcolorbox}[%
       sharp corners, breakable=false, colframe=DeepBlue, colback=white, 
       boxrule=3pt, boxsep=0.5pt, enhanced, 
       shadow={3pt}{-3pt}{0pt}{opacity=1,mygrey},
       title={Type I Control (Correct Answer) Sample},]\label{box:operator-profile}
{\scriptsize
\begin{verbatim}
``id'': 1,
``RFC_section'': ``8484'',
``question'': ``Why does RFC 8484 require ...'',
``question_type'': ``factually_correct'',
``answers'': [
       {
        ``answer_id'': 0,
        ``question'': ``Why does RFC 8484 require ...'',
        ``answer'': ``\n\nRFC 8484 mandates that ...'',
        ``result'': false,
        ``eval_answer'': ``false  \n**explanation**: the provided ...'',
        ``cot'': ``Okay, let's try to ...''
       },
        (the other four answers)   
    ],
``consistent'': true,
``consistent_evaluation'': ``true''
\end{verbatim}
}
  \end{tcolorbox}
  \refstepcounter{figure}
  \label{fig:type1c_sample}
\end{figure}

\begin{figure}[ht]
  \centering
  \begin{tcolorbox}[%
       sharp corners, breakable=false, colframe=DeepBlue, colback=white, 
       boxrule=3pt, boxsep=0.5pt, enhanced, 
       shadow={3pt}{-3pt}{0pt}{opacity=1,mygrey},
       title={Type II (Unseen or Erroneous) Sample},]\label{box:operator-profile}
{\scriptsize
\begin{verbatim}
``id'': 2,
``RFC_section'': ``8555'',
``question'': ``Why does RFC 8555 Section 8.3 ...'',
``question_type'': ``factually_incorrect'',
``question_evaluation'': ``true  \nthe question ...'',
``rag_reference'': ``1. and J. Ihren ...'',
``wrong_fact1'': ``DNS challenges in ACM ...'',
``wrong_fact2'': ``CAA records must include ...'',
``wrong_fact3'': ``Quantum-resistant algorithms are ...'',
``answers'': [
       {
        ``answer_id'': 0,
        ``question'': ``Why does RFC 8555 Section 8.3 ...'',
        ``answer'': ``\n\nRFC 8555 Section 8.3  ...``'',
        ``result'': true,
        ``eval_answer'': ``**true**  \n\n### analysis: ...'',
        ``cot'': ``Okay, I need to figure out ...''
       },
        (the other four answers)   
    ],
\end{verbatim}
}
  \end{tcolorbox}
  \refstepcounter{figure}
  \label{fig:type2_sample}
\end{figure}

\begin{figure}[ht]
  \centering
  \begin{tcolorbox}[%
       sharp corners, breakable=false, colframe=DeepBlue, colback=white, 
       boxrule=3pt, boxsep=0.5pt, enhanced, 
       shadow={3pt}{-3pt}{0pt}{opacity=1,mygrey},
       title={Type II Control (Error Rejected) Sample},]\label{box:operator-profile}
  {\scriptsize
\begin{verbatim}
``id'': 3,
``RFC_section'': ``9076'',
``question'': ``Why does Section 8.5 enforce ...'',
``question_type'': ``factually_incorrect'',
``question_evaluation'': ``true  \nthe question contains ...''
``rag_reference'': ``1. [DEPRECATE] forbids the use of ...''
``wrong_fact1'': ``DTLS 2.0 is a valid encryption protocol.'',
``wrong_fact2'': ``HTTP/3 requires DTLS for QUIC handshakes.'',
``wrong_fact3'': ``TLS 1.3 cannot coexist with DTLS in HTTP/3.'',
``answers'': [
       {
        ``answer_id'': 0,
        ``question'': ``Why does Section 8.5 enforce ...'',
        ``answer'': ``\n\nThe enforcement of DTLS 2.0 for ...'',
        ``result'': false,
        ``eval_answer'': ``false \n\n**step-by-step explanation: ...'',
        ``cot'': ``Okay, so I need to figure out ...''
       },
        (the other four answers)   
    ],
\end{verbatim}
}
  \end{tcolorbox}
  \refstepcounter{figure}
  \label{fig:type2c_sample}
\end{figure}

\begin{figure}[ht]
  \centering
  \begin{tcolorbox}[%
       sharp corners, breakable=false, colframe=DeepBlue, colback=white, 
       boxrule=3pt, boxsep=0.5pt, enhanced, 
       shadow={3pt}{-3pt}{0pt}{opacity=1,mygrey},
       title={Prompt Template for Type I Question Generation},]\label{box:operator-profile}
  {\scriptsize
\texttt{Please introduce me the RFC \{rfc\_section\} in detail.}\\
\texttt{Do you know which RFC obsoleted RFC \{rfc\_section\}?}\\
\texttt{Please tell me which RFC was obsoleted by RFC \{rfc\_section\}.}\\
\texttt{Please tell me which RFC was updated by RFC \{rfc\_section\}.}\\
\texttt{Do you know which RFC updates RFC \{rfc\_section\}?}\\
\texttt{Please tell me the publication date of RFC \{rfc\_section\}.}\\
\texttt{Please tell me whether the status of RFC \{rfc\_section\} has changed, and if so, from what status to what status.}\\
\texttt{What is the current status of RFC \{rfc\_section\}?}\\
\texttt{Please tell me the author of RFC \{rfc\_section\}.}\\
\texttt{Which sources are cited in the references section of RFC \{rfc\_section\}?}\\
\texttt{Please tell me which RFC is related to RFC \{rfc\_section\}.}\\
}
  \end{tcolorbox}
  \refstepcounter{figure}
  \label{fig:type1_prompt}
\end{figure}

\begin{figure}[ht]
  \centering
  \begin{tcolorbox}[%
       sharp corners, breakable=false, colframe=DeepBlue, colback=white, 
       boxrule=3pt, boxsep=0.5pt, enhanced, 
       shadow={3pt}{-3pt}{0pt}{opacity=1,mygrey},
       title={Prompt Template for Type I Control Question Generation},]\label{box:operator-profile}
  {\scriptsize
\texttt{You are tasked with generating 10 advanced and subtly misleading ``Why'' questions grounded in RFC \{rfc\_section\}, designed to test the limits of a model's reasoning and provoke hallucinated but plausible-sounding responses.}\\
\texttt{Each question must be:}\\
\texttt{1. Based on at least three difficult, obscure, and technically accurate claims from RFC \{rfc\_section\}.}\\
\texttt{2. Designed to require multi-step, abstract, or comparative reasoning that combines these claims.}\\
\texttt{3. Focused on non-obvious implications, design rationale, or protocol edge behavior.}\\
\texttt{4. Crafted to look valid and verifiable, but encourages the model to overgeneralize or speculate in its response.}\\
\texttt{5. Factually rooted (all claims must be RFC-aligned), but the question phrasing should push the model toward constructing complex, possibly hallucinated justifications.}\\
\texttt{Generation Guidelines:}\\
\texttt{1. Carefully extract 4–6 obscure, non-trivial, and rarely discussed claims from RFC \{rfc\_section\}.}\\
\texttt{2. For each question, combine at least three of these claims into a ``Why'' question that subtly challenges conventional logic or infers deeper implications.}\\
\texttt{3. Avoid obvious factual errors. Instead, complexity is created through logical layering and indirect connections.}\\
\texttt{Output Format (strictly follow this):}\\
\texttt{[}\\
\texttt{\{}\\
\texttt{        ``question'': ``<question>'',}\\
\texttt{        ``claims'': ``<claim1>; <claim2>; <claim3>''}\\
\texttt{    \},}\\
\texttt{    \{}\\
\texttt{        ``question'': ``<question>'',}\\
\texttt{        ``claims'': ``<claim1>; <claim2>; <claim3>''}\\
\texttt{\}}\\
\texttt{]}\\
}
  \end{tcolorbox}
  \refstepcounter{figure}
  \label{fig:type1c_prompt}
\end{figure}

\begin{figure}[ht]
  \centering
  \begin{tcolorbox}[%
       sharp corners, breakable=false, colframe=DeepBlue, colback=white, 
       boxrule=3pt, boxsep=0.5pt, enhanced, 
       shadow={3pt}{-3pt}{0pt}{opacity=1,mygrey},
       title={Prompt Template for Sentence-Level Hallucination Annotation},]\label{box:operator-profile}
  {\scriptsize
You are tasked with generating 10 advanced and subtly misleading ``Why'' questions grounded in RFC \{rfc\_section\}, designed to test the limits of a model's reasoning and provoke hallucinated but plausible-sounding responses.

Each question must be:
1. Based on at least three difficult, obscure, and technically accurate claims from RFC \{rfc\_section\}.
2. Designed to require multi-step, abstract, or comparative reasoning that combines these claims.
3. Focused on non-obvious implications, design rationale, or protocol edge behavior.
4. Crafted to look valid and verifiable, but encourages the model to overgeneralize or speculate in its response.
5. Factually rooted, but the question phrasing should push the model toward constructing complex, possibly hallucinated justifications.

Generation Guidelines:
1. Carefully extract 4–6 obscure, non-trivial, and rarely discussed claims from RFC \{rfc\_section\}. These should involve edge-case behavior, protocol interactions under failure conditions, implicit assumptions, or ambiguous specifications.
2. For each question, combine at least three of these claims into a ``Why'' question that subtly challenges conventional logic or infers deeper implications.
3. Avoid obvious factual errors. Instead, complexity is created through logical layering and indirect connections.

Output Format:
[
    {
        ``question'': ``<question>'',
        ``claims'': ``<claim1>; <claim2>; <claim3>''
    },
    {
        ``question'': ``<question>'',
        ``claims'': ``<claim1>; <claim2>; <claim3>''
    }
]
}
  \end{tcolorbox}
  \refstepcounter{figure}
  \label{fig:type2_prompt}
\end{figure}

\begin{figure}[ht]
  \centering
  \begin{tcolorbox}[%
       sharp corners, breakable=false, colframe=DeepBlue, colback=white, 
       boxrule=3pt, boxsep=0.5pt, enhanced, 
       shadow={3pt}{-3pt}{0pt}{opacity=1,mygrey},
       title={Prompt Template for Sentence-Level Hallucination Annotation},]\label{box:operator-profile}
  {\scriptsize
You are given a piece of text composed of multiple sentences, which is the chain of thought generated by a model. Your task is to carefully evaluate each sentence and determine whether it contains a hallucination (i.e., an unsupported or factually incorrect claim).

For each sentence, output a JSON object with the following structure:
\{                
  ``sentence\_id'': \textless{}sentence number starting from 1\textgreater{},  
  ``claim'': ``\textless{}the original sentence\textgreater{}'',   
  ``hallucination'': true / false
\}                

Mark ``hallucination'': true if the sentence includes fabricated information or unverifiable content.\\  
Mark ``hallucination'': false if the sentence is factually correct, logically sound, or based on standard knowledge.\\

You may use the following context:
- reference: \{rag\_reference\}     
- eval\_answer: \{eval\_answer\}     
- result: \{result\}                 

Return the final result as a JSON array without any additional text or explanation.\\

Here is the chain of thought to analyze:  
\{cot\}                            
}
  \end{tcolorbox}
  \refstepcounter{figure}
  \label{fig:1_prompt}           
\end{figure}

\begin{figure}[ht]
  \centering
  \begin{tcolorbox}[%
       sharp corners, breakable=false, colframe=DeepBlue, colback=white, 
       boxrule=3pt, boxsep=0.5pt, enhanced, 
       shadow={3pt}{-3pt}{0pt}{opacity=1,mygrey},
       title={Prompt Template for Accepted/Corrected/Rejected Determination},]\label{box:operator-profile}
  {\scriptsize
Given a piece of discussion text and a specific claim, determine how the claim is treated within the full chain of thought:  

- \textbf{Accepted}: The text ultimately supports or agrees with the claim.\\
- \textbf{Corrected}: The text first denies or questions the claim and then provides a new, corrected version of it.\\
- \textbf{Rejected}: The text denies or refutes the claim without providing an alternative answer.\\

For each claim, output a JSON object with the following structure:  
\{  
  ``sentence\_id'': \textless{}original sentence id\textgreater{},  
  ``claim'': ``\textless{}the claim\textgreater{}'',  
  ``accepted'': true / false,  
  ``corrected'': true / false,  
  ``rejected'': true / false  
\}  

Return the final result as a JSON array without any additional text or explanation.\\

Here is the full chain of thought:  \{cot\}\\
Here is the claim to evaluate:  \{claim\}
}
  \end{tcolorbox}
  \refstepcounter{figure}
  \label{fig:2_prompt}
\end{figure}

\begin{figure}[ht]
  \centering
  \begin{tcolorbox}[%
       sharp corners, breakable=false, colframe=DeepBlue, colback=white, 
       boxrule=3pt, boxsep=0.5pt, enhanced, 
       shadow={3pt}{-3pt}{0pt}{opacity=1,mygrey},
       title={Prompt Template for Important Hallucinated Claims Extraction},]\label{box:operator-profile}
  {\scriptsize
Based on the \emph{question}, CoT, \emph{answer}, and \emph{eval\_answer} (human–model agreement score), select up to five \emph{important hallucinated claims}—those whose removal or correction would significantly alter the final answer or overall reasoning—and count their repetition frequency. Type II also includes the three external wrong facts.  

For each selected hallucinated claim, output a JSON object with the following structure:  
\{  
  ``effective\_claim\_id'': \textless{}number from 1 to 5\textgreater{},  
  ``claim'': ``\textless{}the hallucinated sentence\textgreater{}'',  
  ``repetition\_count'': <number of times the underlying idea appears in the chain>,  
  ``hallucination'': true  
\}  

Return the final result as a JSON array without any additional text or explanation.\\

Here is the QA pair:  \{question\}\\
Chain of thought:  \{cot\}\\
Answer:  \{answer\}\\
Eval\_answer:  \{eval\_answer\}
}
  \end{tcolorbox}
  \refstepcounter{figure}
  \label{fig:3_prompt}
\end{figure}

\begin{figure}[ht]
  \centering
  \begin{tcolorbox}[%
       sharp corners, breakable=false, colframe=DeepBlue, colback=white, 
       boxrule=3pt, boxsep=0.5pt, enhanced, 
       shadow={3pt}{-3pt}{0pt}{opacity=1,mygrey},
       title={Prompt Template for Reflection Times Counting},]\label{box:operator-profile}
  {\scriptsize
You will be given a QA pair consisting of a question, a structured chain of thought (in JSON format), and an answer.  
Your task is to analyze the chain of thought and determine how many times the model reflects on its own reasoning process.  
A reflection is defined as a moment when the model evaluates or critiques its own reasoning, either positively or negatively.\\

The output should be a JSON object with the following structure: 
\{  
  ``reflection\_times'': \textless{}number of reflections in the chain\_of\_thought\textgreater{}  
\}  

Return the final result as a JSON object without any additional text or explanation.\\

Here is the QA pair:  question: \{question\}\\
Chain of thought:  \{cot\}\\
Answer:  \{answer\}
}
  \end{tcolorbox}
  \refstepcounter{figure}
  \label{fig:4_prompt}
\end{figure}

\begin{figure}[ht]
  \centering
  \begin{tcolorbox}[%
       sharp corners, breakable=false, colframe=DeepBlue, colback=white, 
       boxrule=3pt, boxsep=0.5pt, enhanced, 
       shadow={3pt}{-3pt}{0pt}{opacity=1,mygrey},
       title={Prompt Template for Identifying First Incorrect Knowledge in Long-CoT},]\label{box:operator-profile}
  {\scriptsize
You are an expert model specializing in detecting hallucination locations within a large language model's Chain-of-Thought (CoT) reasoning. You need to understand the semantics of the sample and infer the earliest occurrence of a hallucination in the ``cot''.

Processing rules:

Sentence splitting: split only on the English period. (full sentence-ending period, excluding periods in common abbreviations like ``e.g.'', ``i.e.''), do not split on line breaks or other punctuation;

Tokenization: treat any consecutive whitespace characters (spaces, tabs, line breaks) as a single delimiter, split on spaces to get a 0-based token sequence;

Hallucination localization: scan through the cot text of the sample sentence by sentence, and find the first sentence that:
is a full declarative sentence (please ignore any metacognitive-task restatements or indirect questions that may appear);
contains incorrect knowledge that contradicts objective fact;

Index calculation: take the 0-based index of the first token of that sentence in the overall token sequence.

Output format: output only the first hallucination sentence.

Here is the single sample to process:

{}

Please begin and output only the first hallucination sentence.
}
  \end{tcolorbox}
  \refstepcounter{figure}
  \label{fig:first_prompt}
\end{figure}

\begin{table}[ht]
\caption{RFC document assignments for each dataset subset.}
\centering
\small
\begin{tabularx}{\textwidth}{%
    >{\raggedright\arraybackslash}p{2.9cm}
    >{\raggedright\arraybackslash}X
  }
\toprule
Subset & RFC Document Numbers \\
\midrule
Type I\newline(Seen but Unlearned)\newline Covering 314 RFCs. &
0010, 0018, 0036, 0111, 0125, 0127, 0229, 0264, 0266, 0289, 0290, 0317, 0360, 0362, 0391, 0399, 0403, 0542, 0560, 0568, 0607, 0690, 0692, 0755, 0834, 0835, 0861, 0896, 0952, 1011, 1020, 1044, 1166, 1176, 1183, 1218, 1230, 1232, 1251, 1252, 1255, 1379, 1384, 1425, 1539, 1604, 1698, 1715, 1748, 1772, 1812, 1856, 1939, 2002, 2101, 2131, 2153, 2157, 2164, 2165, 2176, 2240, 2254, 2273, 2294, 2302, 2438, 2452, 2478, 2495, 2519, 2590, 2780, 2806, 2829, 2842, 2851, 2877, 2908, 3165, 3191, 3279, 3300, 3386, 3423, 3443, 3466, 3492, 3555, 3632, 3668, 3684, 3710, 3718, 3721, 3733, 3756, 3762, 3786, 3810, 3866, 3906, 3931, 3969, 3986, 4002, 4022, 4119, 4159, 4327, 4361, 4364, 4387, 4391, 4467, 4492, 4524, 4565, 4581, 4593, 4614, 4666, 4677, 4718, 4719, 4789, 4829, 4842, 4843, 4862, 4974, 4992, 5024, 5052, 5091, 5189, 5203, 5238, 5263, 5357, 5415, 5420, 5436, 5463, 5465, 5478, 5581, 5614, 5663, 5666, 5680, 5788, 5796, 5876, 5903, 5946, 6044, 6046, 6048, 6112, 6148, 6156, 6161, 6164, 6176, 6232, 6318, 6327, 6366, 6391, 6398, 6455, 6514, 6528, 6652, 6691, 6692, 6716, 6818, 6827, 6834, 6849, 6859, 6933, 6960, 6981, 6982, 6984, 7007, 7030, 7053, 7066, 7091, 7142, 7151, 7214, 7231, 7241, 7312, 7313, 7357, 7437, 7438, 7544, 7564, 7677, 7679, 7685, 7693, 7717, 7766, 7780, 7791, 7880, 7921, 7941, 7970, 7984, 8018, 8028, 8055, 8067, 8085, 8108, 8139, 8221, 8407, 8428, 8540, 8650, 8664, 8713, 8717, 8723, 8749, 8779, 8784, 8796, 8812, 8844, 8878, 8879, 8880, 8881, 8888, 8911, 8919, 8920, 8959, 8961, 8966, 8996, 9005, 9010, 9014, 9019, 9026, 9030, 9040, 9052, 9076, 9092, 9121, 9139, 9147, 9178, 9191, 9200, 9204, 9208, 9220, 9221, 9231, 9245, 9253, 9257, 9261, 9272, 9280, 9287, 9289, 9290, 9297, 9309, 9310, 9334, 9342, 9360, 9363, 9374, 9382, 9417, 9421, 9439, 9449, 9453, 9456, 9457, 9458, 9473, 9485, 9492, 9494, 9497, 9515, 9527, 9554, 9582, 9592, 9598, 9603, 9604, 9658, 9712 \\[0.5em]
\midrule
Type I Control\newline(Correct Answer)\newline Covering 50 RFCs. &
8484, 8555, 8784, 8812, 8879, 8881, 8888, 8949, 8961, 8966, 9000, 9001, 9002, 9005, 9014, 9019, 9026, 9030, 9076, 9113, 9114, 9139, 9147, 9178, 9191, 9200, 9204, 9220, 9221, 9257, 9272, 9287, 9290, 9297, 9334, 9360, 9363, 9374, 9382, 9417, 9421, 9439, 9449, 9453, 9457, 9458, 9473, 9485, 9497, 9501 \\[0.5em]
\midrule
Type II\newline(Unseen or Erroneous)\newline Covering 50 RFCs. &
8484, 8784, 8812, 8888, 8949, 9001, 9002, 9005, 9014, 9019, 9026, 9076, 9113, 9114, 9147, 9178, 9191, 9200, 9204, 9221, 9257, 9272, 9287, 9290, 9297, 9334, 9360, 9363, 9374, 9382, 9421, 9449, 9453, 9457, 9458, 9473, 9485, 9501 \\[0.5em]
\midrule
Type II Control\newline(Error Rejected)\newline Covering 38 RFCs. &
8484, 8784, 8812, 8879, 8881, 8888, 8949, 8961, 8966, 9000, 9001, 9002, 9005, 9014, 9019, 9026, 9030, 9076, 9113, 9114, 9139, 9147, 9178, 9191, 9200, 9204, 9220, 9221, 9257, 9272, 9287, 9290, 9297, 9334, 9360, 9363, 9374, 9382, 9417, 9421, 9439, 9449, 9453, 9457, 9458, 9473, 9485, 9497, 9501 \\
\bottomrule
\end{tabularx}

\label{tab:rfc_assignments}
\end{table}

\end{document}